\documentclass[12pt]{article}
\usepackage{amssymb, amsmath, bm, float, epsfig, color, slashed, subcaption, multirow, arydshln}

\usepackage[centering]{geometry}
\usepackage{cite}

\allowdisplaybreaks[4]

\makeatletter
\@addtoreset{equation}{section}
\makeatother

\begin{document}

\thispagestyle{empty}

\begin{flushright}
KOBE-TH-20-06 \\
DESY 20-178 \\
\end{flushright}
\vspace{40pt}
\begin{center}
{\Large\bf Index theorem on  $T^2/\mathbb{Z}_N$ orbifolds} \\

\vspace{40pt}
{\bf{Makoto Sakamoto$^\dagger\hspace{.5pt}$}\footnote{E-mail: dragon@kobe-u.ac.jp}}, \, 
{\bf{Maki Takeuchi$^\dagger\hspace{.5pt}$}\footnote{E-mail: 191s107s@stu.kobe-u.ac.jp}}, \, 
{\bf{Yoshiyuki Tatsuta$^{\ast, \, \ddag}\hspace{.5pt}$}\footnote{E-mail: yoshiyuki.tatsuta@sns.it}} \\

\vspace{40pt}
{\it $^\dagger$ Department of Physics, Kobe University, 657-8501 Kobe, Japan \\[5pt]
	$^\ast$ Scuola Normale Superiore and INFN, Piazza dei Cavalieri 7, 56126 Pisa, Italy\\[5pt]
	$\ddag$ Deutsches Elektronen-Synchrotron DESY, 22607 Hamburg, Germany} \\
\end{center}
\vspace{30pt}
%
\begin{abstract}
	\noindent
We investigate chiral zero modes and winding numbers at fixed points on $T^2/\mathbb{Z}_N$ orbifolds. 
It is shown that the Atiyah-Singer index theorem for the chiral zero modes leads to a formula $n_+-n_-=(-V_++V_-)/2N$, where $n_{\pm}$ are the numbers of the $\pm$ chiral zero modes and $V_{\pm}$ are the sums of the winding numbers at the fixed points on $T^2/\mathbb{Z}_N$.
This formula is complementary to our zero-mode counting formula on the magnetized orbifolds with non-zero flux background $M \neq 0$, consistently with substituting $M = 0$ for the counting formula $n_+ - n_- = (2M - V_+ + V_-)/2N$.
\end{abstract}

\newpage
\setcounter{page}{2}
\setcounter{footnote}{0}

\section{Introduction}
%
Superstring theory is known as a unique candidate of the unified theory between gauge interactions and quantum gravity.
In its formulation, the theory requires the presence of extra dimensions due to conformal anomaly cancellation.
A key issue in the long history has been to show that the theory can involve the Standard Model (SM) of particle physics.
Indeed in the context of string phenomenology and string cosmology, many frameworks has been used to construct phenomenological models, e.g., heterotic strings \cite{Kobayashi:2004ya, Buchmuller:2005jr, Buchmuller:2006ik, Lebedev:2006kn, Lebedev:2007hv}, type I setups \cite{Bachas:1995ik, Blumenhagen:2000wh, Angelantonj:2000hi}, type IIA/B setups \cite{Aldazabal:1998mr, Ibanez:2001nd, Cremades:2003qj, Cremades:2004wa, Marchesano:2004yq, Marchesano:2004xz} and F-theory \cite{Vafa:1996xn, Heckman:2008qa, Cecotti:2009zf}.

In higher dimensions, irreducible representations of spinors are vector-like in the unit of four-dimensional (4d) Weyl spinors.
In the stringy frameworks above as well as higher-dimensional model constructions, a crucial difficulty is to obtain chiral spectra like the SM quarks and leptons.
As found in the literature, there are powerful mechanisms to realize the chirality in 4d spacetime: orbifold projections \cite{Dixon:1985jw, Dixon:1986jc}, background magnetic fluxes and Wilson lines \cite{Scherk:1979zr, Witten:1984dg, Bachas:1995ik, Abouelsaood:1986gd} and both \cite{Braun:2006se, Abe:2008fi, Abe:2013bca}.
Phenomenological models including compact extra dimensions have been widely 
expected to solve the problems behind the SM.
In terms of the mechanisms above, the higher-dimensional models are found to lead to three-generation models \cite{Libanov:2000uf, Frere:2000dc, Neronov:2001qv, Aguilar:2006sz, Gogberashvili:2007gg, Guo:2008ia, Abe:2008sx, Kaplan:2011vz, Abe:2015yva}, the quark/charged-lepton mass hierarchy \cite{ArkaniHamed:1999dc,Dvali:2000ha,Gherghetta:2000qt,Kaplan:2000av,Kaplan:2001ga,Huber:2000ie, Fujimoto:2012wv, Fujimoto:2017lln, Abe:2014vza}, CP violation \cite{Fujimoto:2013ki, Kobayashi:2016qag} and so on.
These mechanisms have played important roles also in model constructions of the superstring theory \cite{Ibanez:2001nd, Cvetic:2001nr, Kobayashi:2004ya, Marchesano:2004yq, Cvetic:2009yh} as well as higher-dimensional grand unified theories, e.g., \cite{Kawamura:2000ev, Hebecker:2001wq, Asaka:2001eh}.

A smart way to discriminate whether a given setup is chiral or not, is to check the index
\begin{gather}
\textrm{Ind} (i\slashed D) \equiv n_+-n_-,
\label{eq1.1}
\end{gather}
where $n_{\pm}$ are the number of $\pm$ chiral zero modes for a Dirac operator $\slashed D$ on extra dimensions. 
This is the notion of the Atiyah-Singer index theorem \cite{Atiyah:1963zz}.
The index is a topological invariant and takes non-zero values if the setup contains lowest-lying states with chirality.
The index theorem was applied to a two-dimensional (2d) torus with background magnetic flux \cite{Witten:1984dg,Green:1987mn},
\begin{gather}
n_+-n_- = \frac{q}{2\pi} \int_{T^2} F = M,
\label{eq1.2}
\end{gather}
where $M$ denotes the flux quanta.
For $M \neq 0$, the index is non-zero and it is easily confirmed that the lowest-lying states are chiral and degenerate (e.g., $n_+ = M$ and $n_- = 0$ for $M>0$) thanks to the presence of magnetic flux.

In our previous paper \cite{Sakamoto:2020pev}, 
we have discovered a zero-mode counting formula on magnetized orbifolds $T^2/\mathbb{Z}_N \,\, (N = 2, 3, 4, 6)$ for $M>0$, 
\begin{gather}
n_+-n_- = \frac{M-V_+}{N}+1,
\label{eq1.3}
\end{gather}
where $V_{+}$ denotes the total winding numbers for positive chirality modes 
at fixed points.\footnote{
To be precise, the formula \eqref{eq1.3} has not been established as the
index theorem.
This is because in \cite{Sakamoto:2020pev} the equality in Eq.\,\eqref{eq1.3}
has been verified by computing $n_{+}-n_{-}$ and $(M-V_{+})/N + 1$
separately and then comparing their values.
Thus, it is not still clear 
what leads to the formula \eqref{eq1.3}.
}
Interestingly, both $M/N$ and $V_{+}/N$ are not integers, in general, 
but the combination $(M - V_{+})/N$ becomes an integer in any pattern.
Thus, the winding numbers at the fixed points are especially important 
for the index on the orbifolds.

One would suppose that on the magnetized orbifolds, the index should be affected by two sources, i.e. the flux background $M \neq 0$ and the orbifold projections. Let us separate Eq.\,\eqref{eq1.3} into the flux-dependent part $M/N$ and independent one $-V_+/N + 1$. This makes us suppose that the former originates from the total flux on the orbifolds, where the fundamental area is $1/N$ as much as that on the torus.
In this paper, we further pursuit what the flux-independent term $-V_+/N + 1$ implies.
Considering the index theorem on $T^2/\mathbb{Z}_N$ for $M=0$, 
we will derive another expression of our zero-mode counting formula
\begin{gather}
n_+-n_- = \frac{1}{2N}(-V_++V_-),
\label{eq1.4}
\end{gather}
and find $V_{+} + V_{-} = 2N$.
Here $V_{\pm}$ denote the total winding numbers for $\pm$ chirality modes 
at fixed points.
These relations keep a consistency with substituting Eq.\,\eqref{eq1.3} for $M = 0$.
In addition, the formula \eqref{eq1.4} is a generic expression 
because its both sides are antisymmetric under the exchange of $\pm$.

We prove the formula \eqref{eq1.4} as the index theorem.
To this end, we use the trace formula
\begin{gather}
\textrm{Ind} (i \slashed{D}) 
= n_{+} - n_{-}
=\lim_{\rho \to \infty} \, \textrm{tr} [\sigma_3 e^{\slashed{D}^2/\rho^2}].
\label{eq1.5}
\end{gather}
%
Then, we find that Eq.\,\eqref{eq1.5} leads to the index formula \eqref{eq1.4}.
Our derivation clearly shows that the index $n_{+}-n_{-}$ on $T^{2}/\mathbb{Z}_{N}$ 
is determined by the winding numbers at the fixed points.
The proof is the main result of this paper.
%

This paper is organized as follows. In Section 2, we start with the Lagrangian of a six-dimensional ($6$d) Weyl spinor on a 2d torus $T^2$. 
In Section 3, we explicitly construct mode functions on orbifolds $T^2/\mathbb{Z}_N$. 
The values of $n_{\pm}$ for each $\mathbb{Z}_N$ parity $\eta$, the Scherk-Schwarz twist
phase $(\alpha_1,\alpha_2)$ and $N$ are computed in Section 3.
In Section 4, we evaluate the trace formula \eqref{eq1.5} by using a complete set
of the mode functions, and
then confirm the relation (\ref{eq1.4}) to the index theorem 
from the viewpoint of winding numbers $V_{\pm}$ in Section 5. 
Section 6 is devoted to conclusion and discussion. In appendix A, we derive a formula used in our discussion.

\section{Six-dimensional Weyl fermion on $T^2$}
%
First, we briefly discuss the mode expansion of a 6d Weyl fermion on a 2d torus $T^2$.

 \subsection{Setup}
%
We start with the Lagrangian of a 6d Weyl fermion on $T^2$:
 \begin{gather}
 {\cal L}_{\rm 6d} = i \bar \Psi \Gamma^{ I} \partial_{ I} \Psi \qquad (\Gamma_7 \Psi =+ \Psi),
 \label{eq2.1}
 \end{gather} 
 where $ I \,\, (=0,1,2,3,5,6)$ is the 6d spacetime index and $\Gamma^0,\Gamma^1,\cdots,\Gamma^6$ denote the 6d gamma matrices satisfying 
\begin{gather}
\{ \Gamma^{ I}, \Gamma^{ J}\}=-2\eta^{ I J} \qquad( I, J=0,1,2,3,5,6),
\label{eq2.2}\\
(\Gamma ^{\rm I})^\dag=
\begin{cases} 
+\Gamma^{I}\quad ( I=0),\\
-\Gamma^{ I}\quad ( I\neq 0),
\end{cases}
\label{eq2.3}
\end{gather}
with $ \eta^{ I  J} ={\rm diag} \,\,(-1,1,1,1,1,1)$.
$\Gamma_7$ is the 6d chiral operator defined by $-\Gamma^0  \Gamma^1 \Gamma^2 \Gamma^3 \Gamma^5 \Gamma^6$.

For our analysis, it is convenient to take the following representation of gamma matrices:
\begin{gather}
\Gamma^{ I}=
\begin{cases} 
\gamma^{\mu}\otimes I_2 \qquad  I=\mu=0,1,2,3,\\
\gamma_5 \otimes i \sigma_1  \qquad  I=5,\\
\gamma_5 \otimes i \sigma_2  \qquad  I=6,
\end{cases}
\label{eq2.4}
\end{gather}
where $\sigma_a \,(a=1,2,3)$ denotes the Pauli matrices, \,$I_2$ is the $2\times 2$
unit matrix and $\gamma_5=i \gamma^0 \gamma^1 \gamma^2 \gamma^3$.
The 6d chiral operator $\Gamma_7$ is defined by $\gamma_5 \otimes \sigma_3$.

With this representation of the 6d gamma matrices,
the 6d Weyl fermion $\Psi(x,z)$ can be decomposed into 4d Weyl left/right-handed fermions $\psi^{(4)}_{\textrm{L/R}} (x)$ as
\begin{gather}
\Psi(x,z)=\sum_n \{\psi^{(4)}_{\textrm{R},n} (x) \otimes \psi^{(2)}_{+,n} (z)+ \psi^{(4)}_{\textrm{L},n} (x) \otimes \psi^{(2)}_{-,n} (z)\},
\label{eq2.5}
\end{gather}
where $x^\mu\,(\mu=0,1,2,3)$ denotes the 4d Minkowski coordinate and 
$z=x_5+ix_6$ is the complex coordinate on the 2d torus $T^2$.
Since the 2d Weyl fermions $\psi ^{(2)}_{\pm,n} (z)$ are chosen as
\begin{gather}
\sigma_3 \psi^{(2)}_{\pm,n}=\pm  \psi^{(2)}_{\pm,n},
\label{eq2.6}
\end{gather}
we can express  $\psi ^{(2)}_{\pm,n} $ as the form 
\begin{gather}
\psi^{(2)}_{+, \hspace{.5pt} n}(z) = 
\begin{pmatrix}
f_{+, \hspace{.5pt} n}(z) \\[3pt] 0
\end{pmatrix}
, \qquad \psi^{(2)}_{-, \hspace{.5pt} n}(z) = 
\begin{pmatrix}
0 \\[3pt] f_{-, \hspace{.5pt} n}(z)
\end{pmatrix},
\label{eq2.7}
\end{gather}
where $n$ labels the Kaluza-Klein (KK) levels.

The 2d torus $T^2$ is defined by the identification
\begin{gather}
z\sim z+1 \sim z+\tau \qquad ( \tau \in \mathbb{C} ,\, \rm{Im} \tau>0)
\label{eq2.8}
\end{gather}
under torus lattice shifts.\footnote{
	Since the compactification scale of the torus is irrelevant for our analysis, we will take a radius of $T^2$ to be 1. The complex parameter $\tau$ of $T^2$  specifies the shape of $T^2$.}
The 2d Weyl fermions $ \psi^{(2)}_{\pm,n}(z) \,\, ({\rm{or}} \,\, f_{\pm,n}(z))$ are required to satisfy the boundary conditions
\begin{gather}
 \psi^{(2)}_{\pm,n}(z+1)=e^{i 2\pi \alpha_1} \,  \psi^{(2)}_{\pm,n}(z),
 \label{eq2.9}\\
 \psi^{(2)}_{\pm,n}(z+\tau)=e^{i 2\pi \alpha_2} \,  \psi^{(2)}_{\pm,n}(z),
\label{eq2.10}
\end{gather}
where $\alpha_j\,(j=1,2)$ corresponds to a  Scherk-Schwarz (SS) twist phase.

 \subsection{Mode functions on $T^2$}
%
The mode functions $f_{\pm,n}(z)$ on $T^2$ are taken as  eigenfunctions of the differential operator $-4\partial_z \partial_{\bar z}\,$, i.e.
 \begin{gather}
 -4\partial_z \partial_{\bar z} f_{+,n} (z)=m_{n}^{2} f_{+,n}(z),
 \label{eq2.11}\\
 -4\partial_z \partial_{\bar z} f_{-,n} (z)=m_{n}^{2} f_{-,n}(z).
 \label{eq2.12}
 \end{gather}
 Here, $\partial_z$ and $\partial_{\bar z}$ are defined by
  \begin{gather}
\partial_z =\frac{1}{2}(\partial_5 -i \partial_6),
\qquad \partial_{\bar z} =\frac{1}{2}(\partial_5 +i \partial_6).
 \label{eq2.13}
 \end{gather}
 We then require that the mode functions obey the so-called supersymmetry relations \cite{Fujimoto:2016rbr,Fujimoto:2016llj}
  \begin{align}
 2\partial_{\bar z} f_{+,n} (z)&=-m_{n} f_{-,n}(z),
 \label{eq2.14}\\
 2\partial_{z} f_{-,n} (z)&=+m_{n} f_{+,n}(z)
 \label{eq2.15}
 \end{align}
 without loss of generality.

 It follows from  Eqs.~\eqref{eq2.11}\,--\,\eqref{eq2.15} that the 6d Dirac equation for $\Psi(x,z)$ reduces to the 4d Dirac equations
    \begin{gather}
  i \gamma^{\mu} \partial_{\mu} \psi^{(4)}_{\textrm{R},n}(x)+m_n  \psi^{(4)}_{\textrm{L},n}(x)=0,
  \label{eq2.16}\\
   i \gamma^{\mu} \partial_{\mu} \psi^{(4)}_{\textrm{L},n}(x)+m_n  \psi^{(4)}_{\textrm{R},n}(x)=0.
  \label{eq2.17}
   \end{gather}
   Thus, the KK mass eigenvalue $m_n$ corresponds to the mass of the 4d Dirac fermion $\psi^{(4)}_{n}=\psi^{(4)}_{\textrm{R},n}+\psi^{(4)}_{\textrm{L},n}$ for $m_n\neq 0$.

   The mode functions satisfying the equations (\ref{eq2.11}), (\ref{eq2.12}) and the boundary conditions  (\ref{eq2.9}), (\ref{eq2.10}) are found as
  \begin{align}
f_{\pm,\bm{n}+\bm{\alpha}}(z)&=A_{\pm,\bm{n}+\bm{\alpha}}u_{\bm{n}+\bm{\alpha}}(z),
 \label{eq2.18}\\
u_{\bm{n}+\bm{\alpha}}(z)&\equiv e^{i 2\pi\{(n_1+\alpha_1)y_1+(n_2+\alpha_2)y_2\}}
\qquad (n_1,n_2 \, \in \, \mathbb{Z}),
\label{eq2.19}
 \end{align}  
 where $A_{\pm,\bm{n}+\bm{\alpha}}$ are normalization constants and $\bm y =(y_1,y_2)$ is the oblique coordinate defined by 
  \begin{gather}
z=y_1+\tau y_2 \qquad (0\leq y_1,y_2<1).
 \label{eq2.20}
 \end{gather}  
 The mass eigenvalue $m_{\bm{n}+\bm{\alpha}}^2$ is then given by
 \begin{gather}
 m_{\bm{n}+\bm{\alpha}}^2=(2\pi)^2 \left[ (n_1+\alpha_1)^2+\left( -\frac{\rm {Re}\tau}{\rm{Im}\tau}(n_1+\alpha_1)+\frac{1}{\rm{Im}\tau}(n_2+\alpha_2)\right)^2 \right] ,
 \label{eq2.21}
\end{gather}  
which comes from
\begin{gather}
4\partial_z \partial_{\bar z}=\left( \frac{\partial}{\partial y_1}\right) ^2+\left( -\frac{\rm {Re}\tau}{\rm{Im}\tau}\frac{\partial}{\partial y_1}+\frac{1}{\rm{Im}\tau}\frac{\partial}{\partial y_2}\right)^2 .
\label{eq2.22}
\end{gather}

It follows from Eq.\,\eqref{eq2.21} that only when $\bm {n}=\bm{ \alpha}=\bm{0}$, there exist the chiral zero-mode solutions such that $m_{\bm0}=0$. This shows
\begin{gather}
n_+=n_-=
\begin{cases}
1 \quad \textrm{for} \,(\alpha_1,\alpha_2)=(0,0)\quad \textrm{mod} \,\, 1,\\
0 \quad \textrm{for} \,(\alpha_1,\alpha_2)\neq(0,0)\quad \textrm{mod} \,\, 1.
\end{cases}
\label{eq2.23}
\end{gather}  
Then the index on $T^2$ is given as
\begin{gather}
 {\rm Ind} \, (i\slashed{D}) \equiv n_{+} - n_{-}=0.
\label{eq2.24}
\end{gather}
Namely, the lowest-lying states are always vector-like. 
Therefore, model constructions on $T^2$ is less interesting from an index theorem point of view.
As we will see, the index for $T^2/\mathbb{Z}_N \, (N=2,3,4,6)$ orbifold setups, however, can be nontrivial due to winding numbers at fixed points on $T^2/\mathbb{Z}_N$. Our main motivation of this paper is to show it.

\section{Mode functions on $T^2/\mathbb{Z}_{N}$ orbifolds}

%
\subsection{$\mathbb{Z}_N$ eigen mode functions}
%
Let us now proceed to  $T^2/\mathbb{Z}_{N}$ orbifolds.
The $T^2/\mathbb{Z}_{N}$ orbifold is defined by the torus identification $(z \sim z+1 \sim z+\tau) $ and an additional $\mathbb{Z}_{N}$ one 
\begin{gather}
z \sim \omega z \qquad(\omega=e^{i2\pi/N}).
\label{eq3.1}
\end{gather}

It has already been known that there exist only four kinds of the orbifolds: $T^2/\mathbb{Z}_{N} \, \,(N=2,3,4,6)$.
For $N=2$, there is no limitation on $\tau$ except for $\rm{Im} \tau >0$. For $N=3,4$ and $6$, $\tau$ must be equivalent to $\omega$ because of crystallography \cite{Choi:2006qh}.
For convenience, we will use both $\tau$ and $\omega$.
It should be noticed that in order to be consistent with the orbifold identification (\ref{eq3.1}), the SS twist phase $(\alpha_1,\alpha_2)$ has to be quantized \cite{Abe:2013bca} such that 
\begin{gather}
(\alpha_1,\alpha_2)=
\begin{cases}
(0,0),(1/2,0),(0,1/2),(1/2,1/2) 
&\qquad \textrm{for} \,T^2/\mathbb{Z}_{2},\\
(0,0),(1/3,1/3),(2/3,2/3)
&\qquad \textrm{for} \,T^2/\mathbb{Z}_{3},\\
(0,0),(1/2,1/2)
& \qquad \textrm{for} \,T^2/\mathbb{Z}_{4},\\
(0,0) 
&\qquad \textrm{for} \,T^2/\mathbb{Z}_{6}.
\end{cases}
\label{eq3.2}
\end{gather}

Mode functions on the  $T^2/\mathbb{Z}_{N}$ orbifold are classified by $\mathbb{Z}_{N}$ eigenvalues $\eta=\omega^k \,\, (k=0,1,\cdots,N-1)$ under the $\mathbb{Z}_N$ rotation $z \to \omega z $ such as
\begin{align}
f_{+,\bm{n}+\bm{\alpha}}(\omega z)&=\eta f_{+,\bm{n}+\bm{\alpha}}(z),
\label{eq3.3}\\
f_{-,\bm{n}+\bm{\alpha}}(\omega z)&=\omega \eta f_{-,\bm{n}+\bm{\alpha}}(z).
\label{eq3.4}
\end{align}
We emphasize that if the $\mathbb{Z}_{N}$ eigenvalue of $f_{+,\bm{n}+\bm{\alpha}}(z)$ is $\eta$, then that of $f_{-,\bm{n}+\bm{\alpha}}(z)$ has to be $\omega \eta$. This additional factor $\omega$ comes from a rotation matrix acting on 2d spinors  \cite{Abe:2013bca}, and is necessary to be compatible with the supersymmetry relations (\ref{eq2.14}) and (\ref{eq2.15}).

In terms of the mode functions on $T^2$, those on $T^2/\mathbb{Z}_{N}$ can be constructed as
\begin{gather}
\xi_{\bm{n}+\bm{\alpha}}^{\eta}(z)=A_{\bm{n}+\bm{\alpha}} \sum_{l=0}^{N-1} \bar{\eta}^l u_{\bm{n}+\bm{\alpha}}(\omega^l z) ,
\label{eq3.5}
\end{gather}
which belongs to the $\mathbb{Z}_N$ eigenvalue $\eta$:
\begin{gather}
\xi_{\bm{n}+\bm{\alpha}}^{\eta}(\omega z)=\eta \, \xi_{\bm{n}+\bm{\alpha}}^{\eta}( z).
\label{eq3.6}
\end{gather}
Here, $A_{\bm{n}+\bm{\alpha}}$ is a normalization constant.

We can show that the mode function $u_{\bm{n}+\bm{\alpha}}(z)$ on $T^2$ satisfies the relations
\begin{align}
u_{\bm{n}+\bm{\alpha}}(\omega^l (z+1))&=e^ {i2\pi\alpha_1}u_{\bm{n}+\bm{\alpha}}(\omega^l z),
\label{eq3.7}\\
u_{\bm{n}+\bm{\alpha}}(\omega^l (z+\tau))&=e^ {i2\pi\alpha_2}u_{\bm{n}+\bm{\alpha}}(\omega^l z)
\label{eq3.8}
\end{align}
for $l=0,1,\cdots,N-1$. Note that  Eqs.~\eqref{eq3.7} and (\ref{eq3.8}) hold only when the SS twist phase $(\alpha_1,\alpha_2)$ is quantized as Eq.\,\eqref{eq3.2}.
From Eqs.~\eqref{eq3.7} and (\ref{eq3.8}) the $\mathbb{Z}_{N}$ eigen mode function $\xi_{\bm{n}+\bm{\alpha}}^{\eta}(z)$ on  $T^2/\mathbb{Z}_{N}$ satisfies the same boundary conditions as the mode functions on $T^2$, i.e.
\begin{gather}
\xi_{\bm{n}+\bm{\alpha}}^{\eta}(z+1)=e^ {i2\pi\alpha_1}\xi_{\bm{n}+\bm{\alpha}}^{\eta}(z),
\label{eq3.9}\\
\xi_{\bm{n}+\bm{\alpha}}^{\eta}(z+\tau)=e^ {i2\pi\alpha_2}\xi_{\bm{n}+\bm{\alpha}}^{\eta}(z).
\label{eq3.10}
\end{gather}

Under the $\mathbb{Z}_{N}$ rotation, the mode function $u_{\bm{n}+\bm{\alpha}}(z)$ satisfies the relation
\begin{gather}
u_{\bm{n}+\bm{\alpha}}(\omega z)= u_{\omega (\bm{n}+\bm{\alpha})}(z),
\label{eq3.11}
\end{gather}
where $\omega (\bm{n}+\bm{\alpha})$ is an abbreviation of the following quantity:
\begin{gather}
\omega(\bm{n}+\bm{\alpha})=
\begin{cases}
(-n_1-\alpha_1,-n_2-\alpha_2) &\qquad \textrm{for} \,T^2/\mathbb{Z}_{2},\\
(n_2+\alpha_2,-n_1-\alpha_1-n_2-\alpha_2) &\qquad \textrm{for} \,T^2/\mathbb{Z}_{3},\\ 
(n_2+\alpha_2,-n_1-\alpha_1) &\qquad \textrm{for} \,T^2/\mathbb{Z}_{4},\\ 
(n_2,-n_1+n_2) &\qquad \textrm{for} \,T^2/\mathbb{Z}_{6}.
\end{cases}
\label{eq3.12}
\end{gather}
Here, notice that $\bm{\alpha}=\bm 0$ for $T^2/\mathbb{Z}_{6}$.

\begin{figure}[!t]
	\centering
	\includegraphics[width=1.0\textwidth]{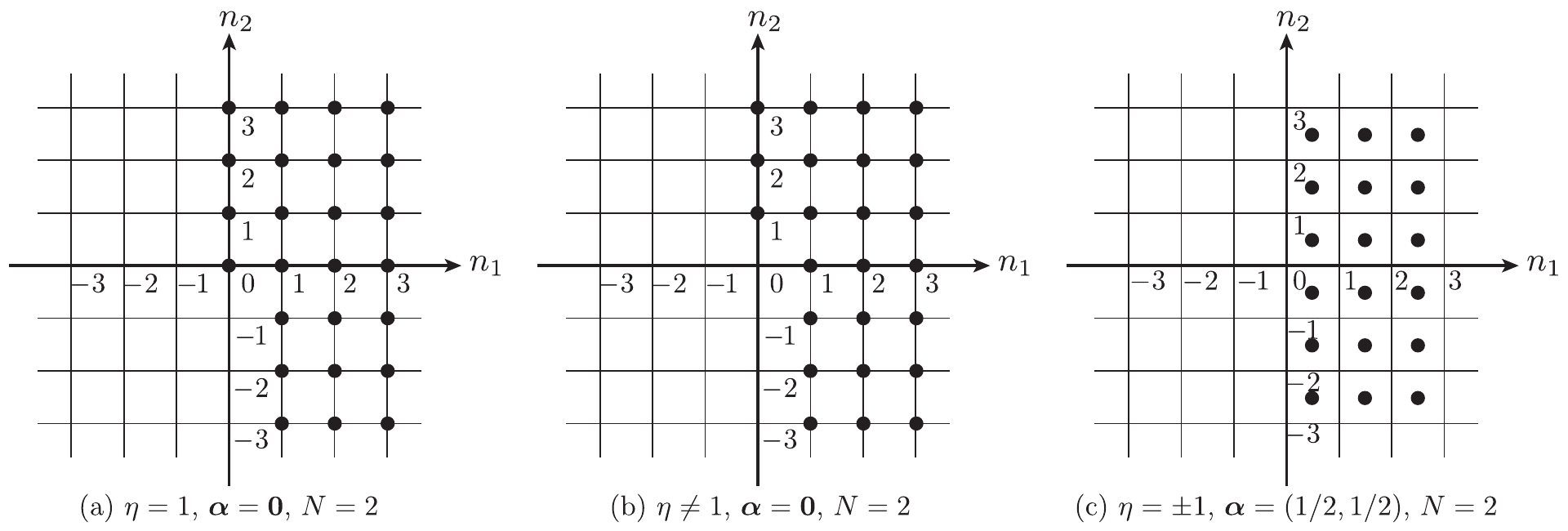}
	\caption{Examples of $\Lambda/\mathbb{Z}_2$. The black dots denote elements belonging to $\Lambda/\mathbb{Z}_2$. }
	\label{figure1}
\end{figure}

Note that $\xi_{\bm{n}+\bm{\alpha}}^{\eta}(z)$ are not always independent of all $n_1,n_2 \in \mathbb Z$.
Since $\xi_{\bm{n+\alpha}}^{\eta}(z)$ obeys the relation
\begin{gather}
\xi_{\omega (\bm{n}+\bm{\alpha})}^{\eta}(z)=\eta \left( \frac{A_{\omega (\bm{n}+\bm{\alpha})}}{A_{\bm{n}+\bm{\alpha}}}\right )\xi_{ \bm{n}+\bm{\alpha}}^{\eta}(z),
\label{eq3.13}
\end{gather}
the independent set of $ \xi_{ \bm{n}+\bm{\alpha}}^{\eta}(z)$ is given by 
\begin{gather}
\left\{ \xi_{\bm{n}+\bm{\alpha}}^{\eta}(z) \  | \  \bm{n}+\bm{\alpha} \in \Lambda / {\mathbb{Z}_{N}} \right\},
\label{eq3.14}
\end{gather}
where
\begin{gather}
\Lambda / {\mathbb{Z}_{N}} =
\begin{cases}
\left\{ \bm{n}+\bm{\alpha} \,\,(n_1,n_2 \in  {\mathbb{Z}}) \,\, | \,\, \bm{n}+\bm{\alpha} \,\, \sim \omega(\bm{n}+\bm{\alpha})  \right\} &\qquad \textrm{for} \,\, \eta=1 \,\, \textrm{or} \,\, \bm{\alpha} \neq \bm 0,\\
\left\{  \bm{n} \,\,(n_1,n_2 \in  {\mathbb{Z}}) \,\, | \,\, \bm{n} \,\, \sim \omega\bm{n} \,\, \textrm{and} \,\, \bm{n}\neq \bm 0 \right\} &\qquad \textrm{for} \,\, \eta \neq 1 \,\, \textrm{and} \,\, \bm{\alpha} = \bm 0,
\end{cases}
\label{eq3.15}
\end{gather}
and
\begin{gather}
\Lambda=
\begin{cases}
\left\{ \bm{n}+\bm{\alpha} \,\,(n_1,n_2 \in  {\mathbb{Z}})  \right\} &\qquad \textrm{for} \,\, \eta=1 \,\, \textrm{or} \,\, \bm{\alpha} \neq \bm 0,\\
\left\{  \bm{n} \,\,(n_1,n_2 \in  {\mathbb{Z}}) \,\, | \,\, \bm{n}\neq \bm{ 0} \right\} &\qquad \textrm{for} \,\, \eta \neq 1 \,\, \textrm{and} \,\, {\bm{\alpha} }= \bm 0.
\end{cases}
\label{eq3.16}
\end{gather}
Notice that ${\bm{n}}=\bm 0$ has to be removed from $\Lambda / {\mathbb{Z}_{N}}$ (and $\Lambda$) for $\eta \neq1$ and ${\bm{\alpha}}=\bm 0$, because $\xi_{\bm 0}^{\eta}(z) $ always vanishes. Explicit examples of $\Lambda / {\mathbb{Z}_{2}}$ are shown in Figure~\ref{figure1}.

The set $\{\xi_{\bm{n}+\bm{\alpha}}^{\eta}(z)\ |\ \bm{n}+\bm{\alpha} \in \Lambda / {\mathbb{Z}_{N}}\}$ of the $\mathbb{Z}_{N}$ eigen modes satisfies
the complete orthonormal condition:
\begin{gather}
\int_{T^2/\mathbb{Z}_{N}}  d^2 z \, (\xi_{\bm{n}+\bm{\alpha}}^{\eta}(z))^{\ast} \, \xi_{\bm{n'}+\bm{\alpha}}^{\eta}(z)
= \delta_{\bm n,\bm n'}
\label{eq3.17}
\end{gather}
with the normalization constant 
\begin{gather}
|A_{\bm{n}+\bm{\alpha}}|^2=
\begin{cases}
(\rm{Im} \tau)^{-1} &\qquad  \textrm{for}  \quad \bm n \neq \bm 0 \quad \textrm{or} \quad \bm{\alpha} \neq \bm 0,\\
(N \rm{Im} \tau)^{-1} &\qquad  \textrm{for}  \quad \bm n =  \bm{\alpha} = \bm 0.
\end{cases}
\label{eq3.18}
\end{gather}
We point out that the normalization constant  (\ref{eq3.18}) is important to derive
Eq.\,(\ref{eq5.5}) in Section 4.

\subsection{Number of zero modes on $T^2/\mathbb{Z}_N$}
%
The mode functions $f_{\pm,\bm{n}+\bm{\alpha}}(z)$ on $T^2/\mathbb{Z}_N$ with the $\mathbb{Z}_N$ transformation properties (\ref{eq3.3}) and (\ref{eq3.4}) are written in terms of $\xi_{\bm{n}+\bm{\alpha}}^{\eta}(z)$ as
\begin{gather}
f_{+,\bm{n}+\bm{\alpha}} (z)=\xi_{\bm{n}+\bm{\alpha}}^{\eta}(z),
\label{eq3.19}\\
f_{-,\bm{n}+\bm{\alpha}}(z)=\xi_{\bm{n}+\bm{\alpha}}^{\omega \eta}(z).
\label{eq3.20}
\end{gather}
The eigenvalue $m_{\bm{n}+\bm{\alpha}}^2$ of $f_{\pm,\bm{n}+\bm{\alpha}} (z)$ is still given by Eq.\,\eqref{eq2.21}. Thus, the chiral zero modes such that $m_{\bm{n}+\bm{\alpha}}=0$ can appear only when $\bm{n}+\bm{\alpha}=\bm 0$ . The lists of the zero modes are summarized in Tables \ref{Z2tab}\,--\,\ref{Z6tab}.

\begin{table}[!t]
	\centering
	{\tabcolsep = 4.5mm
		\renewcommand{\arraystretch}{1.2}
		\scalebox{0.85}{
			\begin{tabular}{cc|c:c|c} \hline
				$\mathbb{Z}_2$ & twist &\multicolumn{2}{c|}{the number of zero modes}  &index \\
				$\eta$  & $(\alpha_1, \alpha_2)$   &\quad \quad$n_+$ \quad\quad&$n_-$&  $n_+ - n_-$ \\ \hline
				$+1$  & $(0,0)$ &$1$ & $0$ &$1$\\
					 & $(\tfrac12,0)$ & $0$ & $0$ & $0$\\
					& $(0,\tfrac12)$ &$0$ & $0$  &$0$\\
						& $(\tfrac12,\tfrac12)$ & $0$ & $0$ & $0$\\\cline{1-5}
					$-1$ & $(0,0)$  &$0$ & $1$ & $-1$\\
						 & $(\tfrac12,0)$ & $0$ & $0$  &$0$\\
						 & $(0,\tfrac12)$  &$0$ & $0$ & $0$\\
						& $(\tfrac12,\tfrac12)$ & $0$ & $0$ & $0$\\\hline
			\end{tabular}
	}
}	
	\caption{The number of the zero modes $f_{\pm,\bm 0}$ on $T^2/\mathbb{Z}_2$ such that $m_{\bm{n}+\bm{\alpha}}=0$.}
	\label{Z2tab}
	\end{table}

\begin{table}[!t]
	\centering
	{\tabcolsep = 4.5mm
		\renewcommand{\arraystretch}{1.2}
		\scalebox{0.85}{
			\begin{tabular}{cc|c:c|c} \hline
				$\mathbb{Z}_3$ & twist &\multicolumn{2}{c|}{the number of zero modes}  &index \\
				$\eta$  & $(\alpha_1, \alpha_2)$ &\quad \quad  $n_+$ \quad \quad  & $n_-$ & $n_+ - n_-$ \\ \hline
				$1$  & $(0,0)$ & $1$ & $0$ & $1$\\
				& $(\tfrac13,\tfrac13)$ & $0$ & $0$ & $0$\\
				& $(\tfrac23,\tfrac23)$ & $0$ & $0$ & $0$\\\cline{1-5}
				$\omega$  & $(0,0)$ & $0$ & $0$ & $0$\\
				& $(\tfrac13,\tfrac13)$ & $0$ & $0$ & $0$\\
				& $(\tfrac23,\tfrac23)$ & $0$ & $0$ & $0$\\\cline{1-5}
				$\omega^2$  & $(0,0)$ & $0$ & $1$ & $-1$\\
				& $(\tfrac13,\tfrac13)$ & $0$ & $0$ & $0$\\
				& $(\tfrac23,\tfrac23)$ & $0$ & $0$ & $0$\\\hline
			\end{tabular}
		}
	}	
	\caption{The number of the zero modes $f_{\pm,\bm 0}$ on $T^2/\mathbb{Z}_3$ such that $m_{\bm{n}+\bm{\alpha}}=0$.}
	\label{Z3tab}
\end{table}

\begin{table}[!t]
	\centering
	{\tabcolsep = 4.5mm
		\renewcommand{\arraystretch}{1.2}
		\scalebox{0.85}{
			\begin{tabular}{cc|c:c|c} \hline
				$\mathbb{Z}_4$ & twist &\multicolumn{2}{c|}{the number of zero modes}  &index \\
				$\eta$  & $(\alpha_1, \alpha_2)$ & \quad \quad$n_+$ \quad \quad& $n_-$ & $n_+ - n_-$ \\ \hline
				$1$  & $(0,0)$ & $1$ & $0$ & $1$\\
				& $(\tfrac12,\tfrac12)$ & $0$ & $0$ & $0$\\\cline{1-5}
					$\omega$  & $(0,0)$ & $0$ & $0$ & $0$\\
				& $(\tfrac12,\tfrac12)$ & $0$ & $0$ & $0$\\\cline{1-5}
					$\omega^2$  & $(0,0)$ & $0$ & $0$ & $0$\\
				& $(\tfrac12,\tfrac12)$ & $0$ & $0$ & $0$\\\cline{1-5}
				$\omega^3$  & $(0,0)$ & $0$ & $1$ & $-1$\\
			& $(\tfrac12,\tfrac12)$ & $0$ & $0$ & $0$\\\hline
			\end{tabular}
		}
	}	
	\caption{The number of the zero modes $f_{\pm,\bm 0}$ on $T^2/\mathbb{Z}_4$ such that $m_{\bm{n}+\bm{\alpha}}=0$.}
	\label{Z4tab}
\end{table}

\begin{table}[!t]
	\centering
	{\tabcolsep = 4.5mm
		\renewcommand{\arraystretch}{1.2}
		\scalebox{0.85}{
			\begin{tabular}{cc|c:c|c} \hline
				$\mathbb{Z}_6$ & twist &\multicolumn{2}{c|}{the number of zero modes}  &index \\
				$\eta$  & $(\alpha_1, \alpha_2)$ & \quad \quad$n_+$ \quad \quad& $n_-$ & $n_+ - n_-$ \\ \hline
				$1$  & $(0,0)$ & $1$ & $0$ & $1$\\\hline
					$\omega$  & $(0,0)$ & $0$ & $0$ & $0$\\\hline
				$\omega^2$  & $(0,0)$ & $0$ & $0$ & $0$\\\hline
					$\omega^3$  & $(0,0)$ & $0$ & $0$ & $0$\\\hline
				$\omega^4$  & $(0,0)$ & $0$ & $0$ & $0$\\\hline
					$\omega^5$  & $(0,0)$ & $0$ & $1$ & $-1$\\\hline
			\end{tabular}
		}
	}	
	\caption{The number of the zero modes $f_{\pm,\bm 0}$ on $T^2/\mathbb{Z}_6$ such that $m_{\bm{n}+\bm{\alpha}}=0$.}
	\label{Z6tab}
\end{table}

From Tables \ref{Z2tab}\,--\,\ref{Z6tab}, we find that the index $n_+ -n_-$ can be non-zero 
and the lowest-lying state in the KK spectrum is chiral.
This property has been used to construct phenomenologically semi-realistic models \cite{Appelquist:2000nn,Dobrescu:2004zi,Burdman:2005sr,Dixon:1985jw,Dixon:1986jc}.
We will prove a nontrivial formula:
\begin{gather}
n_+ -n_-=\frac{1}{2N} (-V_+ + V_-),
\label{eq3.21}
\end{gather}
as an index theorem.
Its nontriviality is that even if $n_+$ and$/$or $n_-$ take zero, 
the sum of the winding numbers $V_{\pm}$ can take non-zero values.
Our derivation clearly shows that the index $n_{+}-n_{-}$ on $T^{2}/\mathbb{Z}_{N}$
can only be determined by the winding numbers at the fixed points, 
as we will see.

\section{Index theorem on  $T^2/\mathbb{Z}_N$ orbifolds}
%

In Sections 4 and 5, we derive the index formula \eqref{eq3.21} by use of the trace formula
\begin{align}
\textrm{Ind} (i \slashed{D}) 
 =\lim_{\rho \to \infty} \, \textrm{tr} [\sigma_3 e^{\slashed{D}^2/\rho^2}]\,. 
\label{eq5.1}
\end{align}
This is our main subject of this paper.
%
%

In terms of the complete orthonormal sets of the mode functions $\{f_{\pm,\bm n +\bm{\alpha}}(z)\}$, the trace  $\lim_{\rho \to \infty} \, \textrm{tr} [\sigma_3 e^{\slashed{D}^2/\rho^2}]$ can be represented as
\begin{align}
&\lim_{\rho \to \infty} \, \textrm{tr} [\sigma_3 e^{\slashed{D}^2/\rho^2}]  \notag \\
&
=\lim_{\rho \to \infty} \, \sum_{\bm n +\bm{\alpha}\, \in \Lambda / {\mathbb{Z}_{N}} }(N_{+,\bm n +\bm{\alpha}}-N_{-,\bm n +\bm{\alpha}}) e^{-m_{\bm n +\bm{\alpha}}^2/\rho^2}\notag \\
&=\lim_{\rho \to \infty} \, \sum_{\bm n +\bm{\alpha}\, \in \Lambda / {\mathbb{Z}_{N}} } \int_{T^2/{\mathbb{Z}_{N}}} d^2 z  \{(\xi_{\bm n +\bm{\alpha}}^{\eta} (z))^{\ast} \xi_{\bm n +\bm{\alpha}}^{\eta} (z)- (\xi_{\bm n +\bm{\alpha}}^{\omega \eta} (z))^{\ast} \xi_{\bm n +\bm{\alpha}}^{\omega \eta} (z)\} e^{-m_{\bm n +\bm{\alpha}}^2/\rho^2},
\label{eq5.2}
\end{align}
where $N_{\pm,\bm n +\bm{\alpha}}$ denote the numbers of the mode functions $f_{\pm,\bm n +\bm{\alpha}}$ and $n_{\pm}\equiv N_{\pm,\bm 0}$. The right-hand-side in the second line of (\ref{eq5.2}) can reduce to $n_+-n_-$ because of the relation $N_{+,\bm n +\bm{\alpha}}=N_{-,\bm n +\bm{\alpha}}$ for $\bm n +\bm{\alpha} \neq \bm 0$.

By using the relation
\begin{gather}
-4\partial_z \partial_{\bar z} \xi_{\bm n +\bm{\alpha}}^{\eta} (z)=m_{\bm n +\bm{\alpha}}^2 \xi_{\bm n +\bm{\alpha}}^{\eta} (z)
\label{eq5.3}
\end{gather}
and a fact that the integration measure $d^2 z$ and $(\xi_{\bm n +\bm{\alpha}}^{\eta(\omega \eta)} (z))^{\ast}   \xi_{\bm n +\bm{\alpha}}^{\eta(\omega \eta)} (z)$ are  invariant under the $\mathbb{Z}_N$ rotation $z \to \omega z$, Eq.\,\eqref{eq5.2} can be rewritten as 
\begin{align}
&\lim_{\rho \to \infty} \, \textrm{tr} [\sigma_3 e^{\slashed{D}^2/\rho^2}] \notag \\
&=\lim_{\rho \to \infty} \, \frac{1}{N} \int_{T^2} d^2 z 
\lim_{z'\to z} e^{4\partial_z \partial_{\bar z}/\rho^2 } 
\hspace{-4mm}
\sum_{\bm n +\bm{\alpha}\, \in \Lambda / {\mathbb{Z}_{N}} } \{ \xi_{\bm n +\bm{\alpha}}^{\eta} (z)(\xi_{\bm n +\bm{\alpha}}^{\eta} (z'))^{\ast}-  \xi_{\bm n +\bm{\alpha}}^{\omega \eta} (z)(\xi_{\bm n +\bm{\alpha}}^{\omega \eta} (z'))^{\ast}\} ,
\label{eq5.4}
\end{align}
where we have replaced $ \int_{T^2/{\mathbb{Z}_{N}}}  d^2 z$ by $({1}/{N}) \int_{T^2} d^2 z$. Inserting the relation (\ref{eq3.5}) into Eq.\,\eqref{eq5.4} and taking Eqs.~\eqref{eq3.15} and (\ref{eq3.18}) into account, we find
\begin{align}
&\lim_{\rho \to \infty} \, \textrm{tr} [\sigma_3 e^{\slashed{D}^2/\rho^2}] \notag \\
&=\lim_{\rho \to \infty} \, \frac{1}{N \rm{Im} \tau} \int_{T^2} d^2 z
\lim_{z'\to z} e^{4\partial_z \partial_{\bar z}/\rho^2 } \sum_{n_1 \in \mathbb{Z}} \sum_{n_2 \in \mathbb{Z}} \sum_{l=0}^{N-1} \eta^l (1-\omega^l) u_{\bm n +\bm{\alpha}}(z)(u_{\bm n +\bm{\alpha}} (\omega^l z'))^{\ast}.
\label{eq5.5}
\end{align}
Eq.\,\eqref{eq5.5} is proved in the appendix.

In the limit of $\rho \to \infty$ and $z'\to z$, the $l=0$ term could diverge like $\delta^2(0)$, but it actually vanishes thanks to the coefficient $(1-\omega^l)$. Therefore, we can take the limit of $\rho \to \infty$ and $z'\to z$ without any divergence or singularity. Then, taking the limit leads to 
\begin{equation}
\lim_{\rho \to \infty} \, \textrm{tr} [\sigma_3 e^{\slashed{D}^2/\rho^2}] 
= \frac{1}{N } \int_{T^2} dy_1 d y_2
\sum_{n_1 \in \mathbb{Z}} \sum_{n_2 \in \mathbb{Z}} \sum_{l=1}^{N-1} \eta^l (1-\omega^l) u_{\bm n +\bm{\alpha}}(z)(u_{\bm n +\bm{\alpha}} (\omega^l z))^{\ast}.
\label{eq5.6}
\end{equation}
Here, we have replaced the integral $\int_{T^2} d^2z$ by ${\rm{Im}} \tau \int_{T^2} dy_1dy_2$, where $\rm{Im} \tau$ corresponds to the area of the 2d torus $T^2$.

\begin{figure}[!t]
	\centering
	\includegraphics[keepaspectratio]{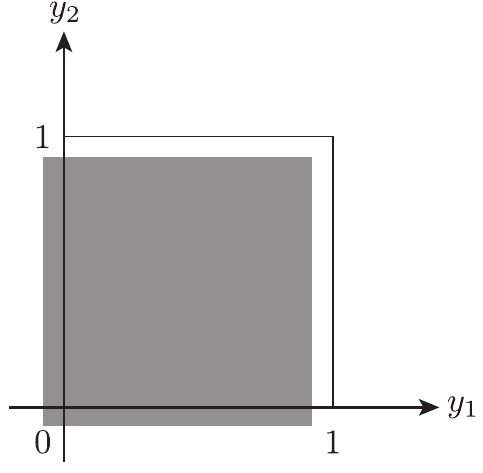}
	\caption{The gray area $(-\varepsilon \leq y_1, y_2 < 1-\varepsilon)$ denotes the fundamental domain of $T^2$ with a small positive number $\varepsilon$.}
	\label{figure3}
\end{figure}

One may take the integral $ \int_{T^2} dy_1 d y_2$ to be $\int_0^1 dy_1 \int_0^1 dy_2$, as usual. However, it is more convenient to choose the fundamental domain of $T^2$, as depicted in Figure \ref{figure3}, in order to avoid troublesome treatment of delta functions appearing on $y_1=0,1$ or $y_2=0,1$.\footnote{Of course, we can reach the same results even if we take the fundamental domain of $T^2$ to be $0\leq y_1,y_2<1$.}

To sum up $n_1$ and $n_2$ in  Eq.\,\eqref{eq5.6}, it is useful to introduce $\bm y^{(l)}=(y_1^{(l)},y_2^{(l)}) \,(l=0,1,\cdots,N-1)$ as
\begin{align}
\omega^l z \equiv y_1^{(l)}+\tau y_2^{(l)} .
\label{eq5.7}
\end{align}
For $l=1$, $(y_1^{(1)},y_2^{(1)})$ is explicitly given by 
\begin{gather}
(y_1^{(1)},y_2^{(1)})=
\begin{cases}
(-y_1,-y_2) \qquad &\textrm{for} \,\,T^2/\mathbb{Z}_2,\\
(-y_2,y_1-y_2) \qquad &\textrm{for} \,\,T^2/\mathbb{Z}_3,\\
(-y_2,y_1) \qquad &\textrm{for} \,\,T^2/\mathbb{Z}_4,\\
(-y_2,y_1+y_2) \qquad &\textrm{for} \,\,T^2/\mathbb{Z}_6.
\end{cases}
\label{eq5.8}
\end{gather}
Then, after summing up $n_1$ and $n_2$ in  Eq.\,\eqref{eq5.6}, the index 
$n_{+}-n_{-}$ is expressed as
\begin{align}
&n_{+}-n_{-} \notag \\
&=\frac{1}{N } \int_{T^2} dy_1 d y_2
\sum_{m_1 \in \mathbb{Z}} \sum_{m_2 \in \mathbb{Z}} \sum_{l=1}^{N-1} \eta^l (1-\omega^l) 
e^{i2\pi (\alpha_1 m_1 + \alpha_2 m_2)} 
\delta(y_1-y_1^{(l)}-m_1)\delta(y_2-y_2^{(l)}-m_2),
\label{eq5.9}
\end{align}
where we have used the formula
\begin{gather}
\sum_{n \in \mathbb{Z}} e^{i2\pi n y} =\sum_{m \in \mathbb{Z}} \delta(y-m).
\label{eq5.10}
\end{gather}
To evaluate Eq.\,\eqref{eq5.9} further,
we will examine $T^2/\mathbb{Z}_N$ orbifolds with $N=2,3,4$ and $6$, separately.

\subsubsection*{\underline{A. Index for  $T^2/\mathbb{Z}_2$}}
%

Let us first discuss the $T^2/\mathbb{Z}_2$ orbifold.
In this case, $(y_1^{(1)},y_2^{(1)})$ is given by $(-y_1,-y_2)$. Inserting it into Eq.\,\eqref{eq5.9} with $N=2$ and $\omega=-1$, we have 
\begin{gather}
n_{+}-n_{-} =\frac{1}{4} \int_{T^2} dy_1 d y_2
\sum_{m_1 \in \mathbb{Z}} \sum_{m_2 \in \mathbb{Z}} \eta
e^{i2\pi (\alpha_1 m_1 + \alpha_2 m_2)} 
\delta(y_1-{m_1}/{2})\delta(y_2-{m_2}/{2}).
\label{eq5.11}
\end{gather}

Since the fundamental domain of $T^2$ has been taken to be $- \varepsilon \leq y_1,y_2 <1- \varepsilon$, the values of $(m_1,m_2)$, which remain in the summation of Eq.\,\eqref{eq5.11} after the $y$-integration, are given by $(m_1,m_2)=(0,0),(1,0),(0,1)$ and $(1,1)$. Then, we find
\begin{align}
n_{+}-n_{-}
&=\frac{1}{4} \int_{T^2} dy_1 d y_2
\left\{ \right.\!  \eta \delta(y_1) \delta(y_2)
  +\eta e^{i2\pi\alpha_1}\delta(y_1-{1}/{2}) \delta(y_2) \notag \\
&\qquad	+\eta e^{i2\pi\alpha_2}\delta(y_1) \delta(y_2-{1}/{2}) 
+\eta e^{i2\pi(\alpha_1+\alpha_2)}\delta\left(y_1-{1}/{2}\right) \delta(y_2-{1}/{2}) \!\left. \right\} \notag \\
&\equiv \frac{1}{4} \int_{T^2} dy_1 d y_2 \sum_{j=1}^{4} W_j \delta^2(\bm y -\bm y_j ^f),
\label{eq5.12}
\end{align}
where $\bm y_j ^f \,(j=1,2,3,4)$ 
are defined by
\begin{gather}
\bm y_1 ^f=(0,0), \quad  \bm y_2 ^f=({1}/{2},0), \quad  \bm y_3 ^f=(0,{1}/{2}), \quad \bm y_4 ^f=({1}/{2},{1}/{2}).
\label{eq5.13} 
\end{gather}

An important observation is that $\bm{y}^{f}_{j}$ given in Eq.\,\eqref{eq5.13}
is just the position of the fixed points on $T^{2}/\mathbb{Z}_{2}$,
as explained below.
Fixed points on $T^{2}/\mathbb{Z}_{N}$ in the complex plane $z = y_{1} + \tau y_{2}$
are defined by
\begin{gather}
z_{f} = \omega z_{f} + m_{1} + m_{2}\tau \qquad \textrm{for}\ 
^{\exists} m_1,m_2\, \in \, \mathbb{Z},
\label{neweq4.1} 
\end{gather}
where $\omega = e^{i2\pi/N}$ for the $T^{2}/\mathbb{Z}_{N}$ orbifold $(N=2,3,4,6)$.
The orbifold fixed points, which are invariant under the $\mathbb{Z}_{N}$
rotation up to torus lattice shifts, are found as
\begin{gather}
z_f=
\begin{cases}
0, 1/2, \tau/2, (1+\tau)/2 &\qquad \textrm{on} \,\,T^2/\mathbb{Z}_2,\\
0, (2+\tau)/3, (1+2\tau)/3 &\qquad \textrm{on} \,\,T^2/\mathbb{Z}_3,\\
0, (1+\tau)/2&\qquad \textrm{on} \,\,T^2/\mathbb{Z}_4,\\
0 &\qquad \textrm{on} \,\,T^2/\mathbb{Z}_6.
\end{cases}
\label{neweq4.2}
\end{gather}
Thus, $\bm{y}^{f}_{j}\ (j=1,2,3,4)$ in Eq.\,\eqref{eq5.13} corresponds
to the position of the fixed points on $T^{2}/\mathbb{Z}_{2}$
in the complex coordinate.
This fact implies that the index $n_{+}-n_{-}$ can only be determined by
information on the fixed points.

The explicit values of $W_{j}\ (j=1,2,3,4)$ are summarized in Table \ref{Z2indextab}.
We can then confirm that the formula \eqref{eq5.12} correctly gives the 
index $n_{+}-n_{-}$ in Table \ref{Z2tab}, as it should be.
However, in the derivation of Eq.\,\eqref{eq5.12}, the physical meaning of $W_{j}$
is less clear.
In the next section, we reveal a geometrical meaning 
of $W_{j}$.


\begin{table}[!t]
\centering
{\tabcolsep = 4.5mm
	\renewcommand{\arraystretch}{1.2}
	\scalebox{0.85}{
			\begin{tabular}{cc|cccc} \hline
				$\mathbb{Z}_2$ & twist &\multicolumn{4}{c}{coefficients 
					of the delta functions}  \\
				$\eta$  & $(\alpha_1, \alpha_2)$ &\, $W_{1}$\, & \,$W_{2}$ \,&\,$W_{3}$ \,&\,$W_{4}$ \, \\ \hline
				$1$  & $(0,0)$ & $1$ & $1$ &$1$ &$1$  \\
				& $(1/2,0)$ & $1$ & $-1$ &$1$ &$-1$  \\
				& $(0,1/2)$ & $1$ & $1$ &$-1$ &$-1$  \\
				& $(1/2,1/2)$ & $1$ & $-1$ &$-1$ &$1$  \\ \cline{1-6}
					$-1$  & $(0,0)$ & $-1$ & $-1$ &$-1$ &$-1$  \\
				& $(1/2,0)$ & $-1$ & $1$ &$-1$ &$1$  \\
				& $(0,1/2)$ & $-1$ & $-1$ &$1$ &$1$  \\
				& $(1/2,1/2)$ & $-1$ & $1$ &$1$ &$-1$  \\ \hline
			\end{tabular}
		}
	}	
	\caption{The coefficients in front of the delta functions in Eq.\,\eqref{eq5.12}.}
	\label{Z2indextab}
\end{table}

\subsubsection*{\underline{B. Index for  $T^2/\mathbb{Z}_3$}}
%

Let us next discuss the index for the $T^2/\mathbb{Z}_3$ orbifold.
In this case, $(y_1^{(l)},y_2^{(l)})\,(l=1,2)$ is given by
\begin{gather}
(y_1^{(1)},y_2^{(1)})=(-y_2,y_1 -y_2), \quad (y_1^{(2)},y_2^{(2)})=(-y_1+y_2,-y_1 ).
\label{eq5.15} 
\end{gather}
Inserting Eq.\,\eqref{eq5.15} into Eq.\,\eqref{eq5.9} with $N=3$, we have 
\begin{align}
&\textrm{Ind} (i\slashed{D}) \notag \\
&=\frac{1}{3} \int_{T^2} dy_1 d y_2
\sum_{m_1 \in \mathbb{Z}} \sum_{m_2 \in \mathbb{Z}} \left\{ \right. \!\eta
(1-\omega)e^{i2\pi (\alpha_1 m_1 + \alpha_2 m_2)} 
\delta(y_1+y_2-m_1)\delta(-y_1 +2y_2-m_2) \notag \\
& \qquad +\eta^2(1-\omega^2)
e^{i2\pi (\alpha_1 m_1 + \alpha_2 m_2)} 
\delta(2y_1-y_2-m_1)\delta(y_2+y_1 -m_2) \!\left.\right\}.
\label{eq5.16}
\end{align}

After the integration of $ \int_{T^2} dy_1 d y_2$, the delta functions $\delta(y_1+y_2-m_1)\delta(-y_1+2y_2-m_2)$ and $\delta(2y_1-y_2-m_1)\delta(y_2+y_1 -m_2)$ in  Eq.\,\eqref{eq5.16} remain only when $(m_1,m_2)=(0,0),(1,0),(1,1)$  and $(m_1,m_2)=(0,0),(1,1),(0,1)$, respectively. Then, it follows that  Eq.\,\eqref{eq5.16} can be written into the form
\begin{gather}
\textrm{Ind} (i\slashed{D}) = \frac{1}{2 \times 3} \int_{T^2} dy_1 d y_2 \sum_{j=1}^{3} W_j \delta^2(\bm y -\bm y_j ^f),
\label{eq5.17}
\end{gather}
where
\begin{align}
W_1&=\frac{2}{3}\left\{\eta (1-\omega)+\eta^2(1-\omega^2)\right\},
\label{eq5.18}\\
W_2&=\frac{2}{3}\left\{\eta (1-\omega)e^{i2\pi \alpha_1}+\eta^2(1-\omega^2)e^{i2\pi(\alpha_1+\alpha_2)}\right\},
\label{eq5.19}\\
W_3&=\frac{2}{3}\left\{\eta (1-\omega)e^{i2\pi (\alpha_1+\alpha_2)}+\eta^2(1-\omega^2)e^{i2\pi\alpha_2}\right\}.
\label{eq5.20}
\end{align}
Here, $\bm y_j ^f \,(j=1,2,3)$ are identified with the position of the fixed points 
on $ T^2/\mathbb{Z}_3$, i.e.
\begin{gather}
\bm y_1 ^f =(0,0), \quad \bm y_2 ^f =({2}/{3},{1}/{3}), \quad
\bm y_3 ^f =({1}/{3},{2}/{3}), \,\,  
\label{eq5.21}
\end{gather}
which correspond to the fixed points given in Eq.\,\eqref{neweq4.2}

The explicit values of $W_{j}\ (j=1,2,3)$ are summarized in Table \ref{Z3indextab}.
We can then confirm that the formula \eqref{eq5.17} correctly gives the
index $n_{+}-n_{-}$ in Table \ref{Z3tab}, as it should be.
In the next section, we reveal the relation between $W_{j}$ and the winding
nmbers at the fixed points \eqref{eq5.21}.

%
%

\begin{table}[!t]
	\centering
	{\tabcolsep = 4.5mm
		\renewcommand{\arraystretch}{1.2}
		\scalebox{0.85}{
			\begin{tabular}{cc|ccc} \hline
				$\mathbb{Z}_3$ & twist &\multicolumn{3}{c}{coefficients of the delta functions}  \\
				$\eta$  & $(\alpha_1, \alpha_2)$ & \quad\,$W_{1}$\quad\, & \quad$W_{2}$ \quad&\quad$W_{3}$ \quad  \\ \hline
				$1$  & $(0,0)$ & $2$ & $2$ &$2$ \\
				& $(1/3,1/3)$ & $2$ & $0$ &$-2$   \\
				& $(2/3,2/3)$ & $2$ & $-2$ &$0$  \\ \cline{1-5}
				$\omega$  & $(0,0)$ & $0$ & $0$ &$0$ \\
				& $(1/3,1/3)$ & $0$ & $-2$ &$2$   \\
				& $(2/3,2/3)$ & $0$ & $2$ &$-2$  \\ \cline{1-5}
				$\omega^2$  & $(0,0)$ & $-2$ & $-2$ &$-2$ \\
				& $(1/3,1/3)$ & $-2$ & $2$ &$0$   \\
				& $(2/3,2/3)$ & $-2$ & $0$ &$2$  \\ \hline
			\end{tabular}
		}
	}	
	\caption{The list of the coefficients in front of the delta functions in Eq.\,\eqref{eq5.17}. }
	\label{Z3indextab}
\end{table}

\subsubsection*{\underline{C. Index for  $T^2/\mathbb{Z}_4$}}
%

Let us discuss the index for the $T^2/\mathbb{Z}_4$ orbifold.
In this case, $(y_1^{(l)},y_2^{(l)})\,\,(l=1,2,3)$ is given by
\begin{gather}
(y_1^{(1)},y_2^{(1)})=(-y_2,y_1), \quad (y_1^{(2)},y_2^{(2)})=(-y_1,-y_2 ),\quad
(y_1^{(3)},y_2^{(3)})=(y_2,-y_1 ).
\label{eq5.23} 
\end{gather}
 Inserting Eq.\,\eqref{eq5.23} into Eq.\,\eqref{eq5.9} with $N=4$, we have 
\begin{align}
\textrm{Ind} (i\slashed{D}) 
&=\frac{1}{4} \int_{T^2} dy_1 d y_2
\sum_{m_1 \in \mathbb{Z}} \sum_{m_2 \in \mathbb{Z}} \notag \\
& \times \left\{\right. \! \eta
(1-\omega)e^{i2\pi (\alpha_1 m_1 + \alpha_2 m_2)} 
\delta(y_1+y_2-m_1)\delta(y_2-y_1-m_2) \notag \\
&+\eta^2(1-\omega^2)
e^{i2\pi (\alpha_1 m_1 + \alpha_2 m_2)} 
\delta(2y_1-m_1)\delta(2y_2-m_2) \notag \\
&+\eta^3(1-\omega^3)
e^{i2\pi (\alpha_1 m_1 + \alpha_2 m_2)} 
\delta(y_1-y_2-m_1)\delta(y_2+y_1-m_2) \! \left.\right\}.
\label{eq5.24}
\end{align}

After the integration of $ \int_{T^2} dy_1 d y_2$, the delta functions $\delta(y_1+y_2-m_1)\delta(y_2-y_1-m_2)$, $\delta(2y_1-m_1)\delta(2y_2 -m_2)$ and
$\delta(y_1-y_2-m_1)\delta(y_2+y_1-m_2)$ in 
 Eq.\,\eqref{eq5.24} remain only when $(m_1,m_2)=(0,0),(1,0)$,  $(m_1,m_2)=(0,0),(1,0),(0,1),(1,1)$ and $(m_1,m_2)=(0,0),(0,1)$, respectively. Then, it follows that  Eq.\,\eqref{eq5.24} can be written into the form
\begin{gather}
\textrm{Ind} (i\slashed{D}) = \frac{1}{2 \times 4} \int_{T^2} dy_1 d y_2 \sum_{j=1}^{4} W_j \delta^2(\bm y -\bm y_j ^f),
\label{eq5.25}
\end{gather}
where
\begin{align}
W_1&=\eta (1-\omega)+\frac{1}{2}\eta^2(1-\omega^2)+\eta^3(1-\omega^3),
\label{eq5.26}\\
W_2&=\eta (1-\omega)e^{i2\pi \alpha_1}+\frac{1}{2}\eta^2(1-\omega^2)e^{i2\pi(\alpha_1+\alpha_2)}+\eta^3(1-\omega^3)e^{i 2\pi \alpha_2},
\label{eq5.27}\\
W_3&=\frac{1}{2}\eta^2(1-\omega^2)e^{i2\pi\alpha_1},
\label{eq5.28} \\
W_4&=\frac{1}{2}\eta^2(1-\omega^2)e^{i2\pi\alpha_2}.
\label{eq5.29}
\end{align}
Here, $\bm y_j ^f \,(j=1,2,3,4)$ in Eq.\,\eqref{eq5.25} are defined by
\begin{gather}
\bm y_1 ^f =(0,0), \quad \bm y_2 ^f =({1}/{2},{1}/{2}), \quad
\bm y_3 ^f =({1}/{2},0), \quad  \bm y_4 ^f =(0,{1}/{2}).
\label{eq5.30}
\end{gather}
The $\bm{y}^{f}_{1}$ and $\bm{y}^{f}_{2}$ correspond to the fixed points on
$T^{2}/\mathbb{Z}_{4}$ given in Eq.\,\eqref{neweq4.2}.
Interestingly, we found additional contributions from the points
$\bm{y}^{f}_{3}$ and $\bm{y}^{f}_{4}$.
Since the $\mathbb{Z}_{4}$ group includes $\mathbb{Z}_{2}$ as its subgroup,
there are additional two ``$\mathbb{Z}_2$ fixed points" that are not invariant 
under the $\mathbb{Z}_4$ rotation but invariant under such a subgroup 
$\mathbb{Z}_2$  $(z \to \omega^2 z = -z)$ up to torus lattice shifts.
Indeed, $\bm{y}^{f}_{3}$ and $\bm{y}^{f}_{4}$ are the ``$\mathbb{Z}_2$ fixed points".

%
The explicit values of $W_j \,(j=1,2,3,4)$ are summarized in Table \ref{Z4indextab}.
We can then confirm that the formula \eqref{eq5.25} correctly gives the index
$n_{+}-n_{-}$ in Table \ref{Z4tab}, as it should be.
In the next section, we show that $W_{j}$ is related to the winding numbers at
the fixed points \eqref{eq5.30}.

%
%

\begin{table}[!t]
	\centering
	{\tabcolsep = 4.5mm
		\renewcommand{\arraystretch}{1.2}
		\scalebox{0.85}{
			\begin{tabular}{cc|cccc} \hline
				$\mathbb{Z}_4$ & twist &\multicolumn{4}{c}{coefficients of the delta functions}  \\
				$\eta$  & $(\alpha_1, \alpha_2)$ &\, $W_{1}$ \,& \,$W_{2}$ \,&\,$W_{3}$ \,&\,$W_{4}$  \\ \hline
				$1$  & $(0,0)$ & $3$ & $3$ &$1$ &$1$ \\
				& $(1/2,1/2)$ & $3$ & $-1$ &$-1$   &$-1$\\ \hline
				$\omega$  & $(0,0)$ & $1$ & $1$ &$-1$ &$-1$ \\
				& $(1/2,1/2)$ & $1$ & $-3$ &$1$   &$1$\\ \hline
		$\omega^2$  & $(0,0)$ & $-1$ & $-1$ &$1$ &$1$ \\
		& $(1/2,1/2)$ & $-1$ & $3$ &$-1$   &$-1$\\ \hline
			$\omega^3$  & $(0,0)$ & $-3$ & $-3$ &$-1$ &$-1$ \\
			& $(1/2,1/2)$ & $-3$ & $1$ &$1$   &$1$\\ \hline
			\end{tabular}
		}
	}	
	\caption{The list of the coefficients $W_j$ in front of the delta functions in Eq.\,\eqref{eq5.25}.}
	\label{Z4indextab}
\end{table}

\subsubsection*{\underline{D. Index  for  $T^2/\mathbb{Z}_6$}}
%

Let us finally discuss the index for the $T^2/\mathbb{Z}_6$ orbifold.
In this case, $(y_1^{(l)},y_2^{(l)})\,(l=1,2,3,4,5)$ is given by
\begin{gather}
(y_1^{(1)},y_2^{(1)})=(-y_2,y_1+y_2), \quad (y_1^{(2)},y_2^{(2)})=(-y_1-y_2,y_1 ),\,\,\notag \\
(y_1^{(3)},y_2^{(3)})=(-y_1,-y_2), \quad
(y_1^{(4)},y_2^{(4)})=(y_2,-y_1-y_2 ),\quad (y_1^{(5)},y_2^{(5)})=(y_1+y_2,-y_1 ).
\label{eq5.32} 
\end{gather}
Inserting Eq.\,\eqref{eq5.32} into Eq.\,\eqref{eq5.9} with $N=6$ and computing in the same way, we arrive at 
\begin{gather}
\textrm{Ind} (i\slashed{D}) = \frac{1}{2 \times 6} \int_{T^2} dy_1 d y_2 \sum_{j=1}^{6} W_j \delta^2(\bm y -\bm y_j ^f),
\label{eq5.33}
\end{gather}
where
\begin{align}
W_1&=2\eta (1-\omega)+\frac{2}{3}\eta^2(1-\omega^2)+\frac{1}{2}\eta^3(1-\omega^3)+\frac{2}{3}\eta^4(1-\omega^4)+2\eta^5(1-\omega^5),
\label{eq5.34}\\
W_2&=W_3=\frac{2}{3}\eta^2(1-\omega^2)+\frac{2}{3}\eta^4(1-\omega^4),
\label{eq5.35}\\
W_4&=W_5=W_6=\frac{1}{2}\eta^3(1-\omega^3).
\label{eq5.36} 
\end{align}
Here, $\bm y_j ^f \,(j=1,2,\cdots,6)$ in Eq.\,\eqref{eq5.33} are defined by
\begin{gather}
\bm y_1 ^f =(0,0), \quad \bm y_2 ^f =({1}/{3},{1}/{3}), \quad
\bm y_3 ^f =({2}/{3},{2}/{3}), \,\, \notag \\ \bm y_4 ^f =({1}/{2},0), \quad
 \bm y_5 ^f =(0,{1}/{2}), \quad \bm y_6 ^f =({1}/{2},{1}/{2}).
\label{eq5.37}
\end{gather}
%
The $\bm{y}^{f}_{1}$ corresponds to a single fixed point on $T^{2}/\mathbb{Z}_{6}$
given in Eq.\,\eqref{neweq4.2}.
Since the $\mathbb{Z}_{6}$ group includes its subgroups $\mathbb{Z}_{3}$
and $\mathbb{Z}_{2}$, there are additional two ``$\mathbb{Z}_3$ fixed points"
and three ``$\mathbb{Z}_2$ fixed points" that are not invariant under 
the $\mathbb{Z}_{6}$ rotation but invariant under such $\mathbb{Z}_{3}$
and $\mathbb{Z}_{2}$ rotations up to torus lattice shifts, respectively.
The two $\mathbb{Z}_{3}$ and three $\mathbb{Z}_{2}$ fixed points are
just given by $\bm{y}^{f}_{2}, \bm{y}^{f}_{3}$
and $\bm{y}^{f}_{4}, \bm{y}^{f}_{5}, \bm{y}^{f}_{6}$ in Eq.\,\eqref{eq5.37},
respectively.

The explicit values of $W_{j}\ (j=1,2,\cdots,6)$ are summarized in Table \ref{Z6indextab}.
We can then confirm that the formula \eqref{eq5.33} correctly gives the index
$n_{+}-n_{-}$ in Table \ref{Z6tab}, as it should be.
In the next section, we show that $W_{j}$ is related to the winding numbers at
the fixed points \eqref{eq5.37}.

%
%

\begin{table}[!t]
	\centering
	{\tabcolsep = 4.5mm
		\renewcommand{\arraystretch}{1.2}
		\scalebox{0.85}{
			\begin{tabular}{c|cccccc} \hline
				$\mathbb{Z}_6$  &\multicolumn{6}{c}{coefficients of the delta functions}  \\
				$\eta$  & $W_{1}$ & $W_{2}$ &$W_{3}$ &$W_{4}$ & $W_{5}$ &$W_{6}$\\ \hline
				$1$  & $5$ & $2$ &$2$ &$1$&$1$ &$1$ \\\hline
			$\omega$  & $3$ & $0$ &$0$ &$-1$&$-1$ &$-1$ \\\hline
			$\omega^2$  & $1$ & $-2$ &$-2$ &$1$&$1$ &$1$ \\\hline
			$\omega^3$  & $-1$ & $2$ &$2$ &$-1$&$-1$ &$-1$ \\\hline
			$\omega^4$  & $-3$ & $0$ &$0$ &$1$&$1$ &$1$ \\\hline
			$\omega^5$  & $-5$ & $-2$ &$-2$ &$-1$&$-1$ &$-1$ \\\hline
			\end{tabular}
		}
	}	
	\caption{The list of the coefficients $W_j$ in front of the delta functions in Eq.\,\eqref{eq5.33}. Here, we omit the SS twist phase because of $\alpha_1=\alpha_2=0$.}
		\label{Z6indextab}
\end{table}

\section{Winding numbers at fixed points on $T^2/\mathbb{Z}_N$}
%

In this section, we compute the winding numbers at fixed points on $T^{2}/\mathbb{Z}_{N}$
and clarify the geometrical meaning of the coefficients $W_{j}$ in front of the
delta functions in Eqs.\,\eqref{eq5.12}, \eqref{eq5.17}, \eqref{eq5.25} and \eqref{eq5.33}.

\begin{figure}[!t]
	\centering
	\includegraphics[width=0.8\textwidth]{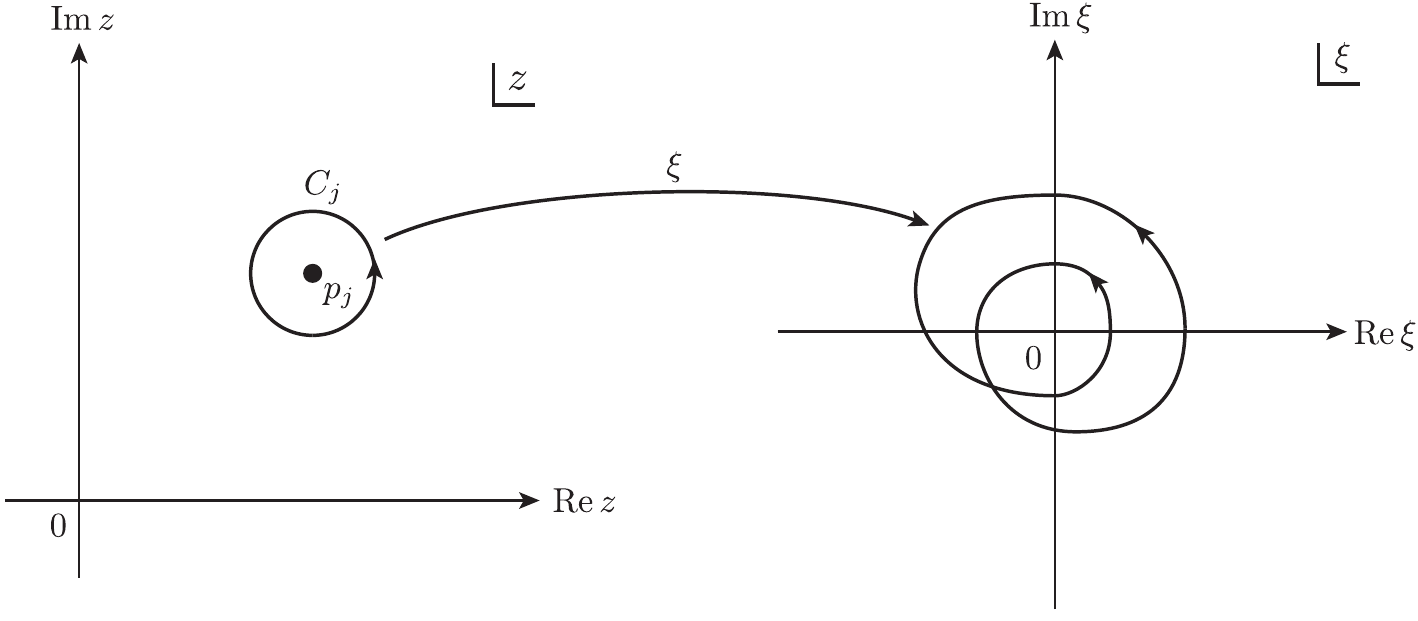}
	\caption{$C_j$ denotes an anti-clockwise contour around the fixed point $z=p_j$. In this example, the winding number is equal to $+2$.}
	\label{figure2}
\end{figure}

Let us define the winding number for the $\mathbb{Z}_N$ eigen modes $\xi_{\bm{n}+\bm{\alpha}}^{\eta}(z)$ as
\begin{gather}
\chi_j (\eta,\bm{\alpha}) \equiv \frac{1}{2\pi i} \oint_{C_j} d{\bm l} \cdot \nabla \log (\xi_{\bm{n}+\bm{\alpha}}^{\eta}(z)),
\label{eq4.3}
\end{gather}
where $C_j$ denotes a sufficiently small circle encircled anti-clockwise around a fixed point $z=p_j$. The line integral along the contour $C_j$ gives a winding number (or occasionally called vortex number), i.e. how many times $\xi_{\bm{n}+\bm{\alpha}}^{\eta}(z)$ wraps around the origin, as illustrated in Figure~\ref{figure2}. Note that if $\xi_{\bm{n}+\bm{\alpha}}^{\eta}(z)$ does not vanish at $z=p_j$, the winding number $\chi_j (\eta,\bm{\alpha})$ obviously takes zero due to  the ``Cauchy  integral formula" in $\xi$ space. The important properties of $\xi_{\bm{n}+\bm{\alpha}}^{\eta}(z)$ in Eq.\,\eqref{eq4.3} are determined only by the $\mathbb{Z}_N$ transformation (\ref{eq3.6}) and the boundary conditions  (\ref{eq3.9}) and  (\ref{eq3.10}).

We are now ready to define the winding numbers for the mode functions $f_{+,\bm{n}+\bm{\alpha}} (z)$ and $f_{-,\bm{n}+\bm{\alpha}} (z)$. Then, we define the winding numbers $\chi_{\pm j}$ for the mode functions $f_{\pm,\bm{n}+\bm{\alpha}} (z)$ around the fixed point $z=p_j$, as\footnote{In the previous paper  \cite{Sakamoto:2020pev}, we have introduced the winding number only for the mode function $f_+$. Here, we need to define that for $f_-$ also in order to show the formula  (\ref{eq3.21}). 
}
\begin{align}
\chi_{+j} &\equiv \chi_j (\eta,\bm{\alpha})  \qquad \textrm{for} \,\, f_{+,\bm{n}+\bm{\alpha}},
\label{eq4.4}\\
\chi_{-j} &\equiv \chi_j (\bar{\omega} \bar{\eta},-\bm{\alpha}) \qquad \textrm{for} \,\, f_{-,\bm{n}+\bm{\alpha}}.
\label{eq4.5}
\end{align}

One might define the winding number $\chi_{-j}$ for $f_{-,\bm{n}+\bm{\alpha}}$
by $\chi_{j}(\omega\eta,\bm{\alpha})$, instead of 
$\chi_j (\bar{\omega} \bar{\eta},-\bm{\alpha})$, since 
$f_{-,\bm{n}+\bm{\alpha}} = \xi_{\bm{n}+\bm{\alpha}}^{\omega\eta}$.
This is not, however, the case.
As we will see later, the definition \eqref{eq4.5} for $\chi_{-j}$
leads to the expected result
\begin{gather}
W_{j} = -\chi_{+j} + \chi_{-j},
\label{neweq4.6}
\end{gather}
otherwise we will not obtain any meaningful relation.

Another reason to adopt the definition \eqref{eq4.5} may be explained as follows. 
To this end, let us consider the 6d charge conjugation $\mathcal{C}$ 
to the 6d fermion $\Psi (x,z)$:
\begin{gather}
\Psi (x,z) \, {\overset{\mathcal{C}}\longrightarrow} \,\Psi^{(\rm{C})} (x,z)=C \bar{\Psi} ^T(x,z).
\label{eq4.6}
\end{gather}
The 6d charge conjugation matrix $C$ is represented as
\begin{gather}
C=C^{(4)} \otimes i\sigma_2,
\label{eq4.7}
\end{gather}
where $C^{(4)}$ is the 4d charge conjugation matrix. Under this charge conjugation, the mode functions $f_{\pm,\bm{n}+\bm{\alpha}}$ transform as\footnote{The transformation  (\ref{eq4.8}) is consistent with the supersymmetry relations  (\ref{eq2.14}) and  (\ref{eq2.15}).}
\begin{gather}
f_{\pm,\bm{n}+\bm{\alpha}}\, {\overset{\mathcal{C}}\longrightarrow} \,f_{\pm,\bm{n}+\bm{\alpha}}^{(\rm{C})} = \pm (f_{\mp,\bm{n}+\bm{\alpha}})^{\ast}.
\label{eq4.8}
\end{gather}
Then $f_{+,\bm{n}+\bm{\alpha}}^{(\rm{C})}(z)=(f_{-,\bm{n}+\bm{\alpha}}(z))^{\ast}$ satisfies
\begin{align}
f_{+,\bm{n}+\bm{\alpha}}^{(\rm{C})}(\omega z)&=\bar{\omega} \bar{\eta}f_{+,\bm{n}+\bm{\alpha}}^{(\rm{C})}(z),
\label{eq4.9}\\
f_{+,\bm{n}+\bm{\alpha}}^{(\rm{C})}( z+1)&=e^{-i2\pi \alpha_1}f_{+,\bm{n}+\bm{\alpha}}^{(\rm{C})}(z),
\label{eq4.10}\\
f_{+,\bm{n}+\bm{\alpha}}^{(\rm{C})}( z+\tau)&=e^{-i2\pi \alpha_2}f_{+,\bm{n}+\bm{\alpha}}^{(\rm{C})}(z).
\label{eq4.11}
\end{align}
The above transformation properties bring another reason to adopt Eq.\,\eqref{eq4.5} as the winding number for $f_{-,\bm{n}+\bm{\alpha}}$.

In the following, we will define the winding numbers $\chi_{\pm j} $ on the fundamental domain of $T^2$ even for  the orbifold $T^2/\mathbb{Z}_N$. If one defines the winding numbers on the fundamental domain of the orbifold $T^2/\mathbb{Z}_N$, instead of $T^2$, the sum of the winding numbers $\chi_{\pm j}$ at fixed points should be divided by $N$, i.e. $\sum_j \chi_{\pm j}/N$ due to $1/N$ reduced area and the deficit angles around the fixed points in comparison with that of the torus.

\subsubsection*{\underline{A. Winding numbers for $T^2/\mathbb{Z}_2$}}
%

In the following, we examine the winding numbers $\chi_{\pm j}$ for the mode functions on  $T^2/\mathbb{Z}_2$ at the fixed points \eqref{eq5.13}, which correspond to
\begin{gather}
p_1=0,\quad p_2=\frac{1}{2},\quad p_3=\frac{\tau}{2},\quad p_4=\frac{1}{2}+\frac{\tau}{2}
\label{eq4.12}
\end{gather}
in the complex plane.

Under the $\mathbb{Z}_2$  rotation $z \, \to \, \omega z \,\,(\omega=-1)$ around the fixed points $p_j \,\, (j=1,2,3,4)$, the $\mathbb{Z}_2$ eigen mode function $\xi_{\bm{n}+\bm{\alpha}}^{\eta}(z)$ is found to satisfy the relations 
\begin{align}
\xi^{\eta}(\omega z)&=\omega^{k} \xi^{\eta} (z),
\label{eq4.13}\\
\xi^{\eta}(\omega z+{1}/{2})&=\omega^{k+2\alpha_1} \xi^{\eta}(z+{1}/{2}),
\label{eq4.14}\\
\xi^{\eta}(\omega z+{\tau}/{2})&=\omega^{k+2\alpha_2} \xi^{\eta} (z+{\tau}/{2}),
\label{eq4.15}\\
\xi^{\eta}(\omega z+{1}/{2}+{\tau}/{2})&=\omega^{k+2\alpha_1+2\alpha_2} 
\xi^{\eta}(z+{1}/{2}+{\tau}/{2}),
\label{eq4.16}
\end{align}
where $\eta=\omega^k \,\,(k=0,1)$. Since the label $\bm n+\bm{\alpha}$ of the mode function $\xi_{\bm{n}+\bm{\alpha}}^{\eta}(z)$ is irrelevant for the analysis below, we omit it, unless otherwise stated.

By plugging $z=0$ into the relations (\ref{eq4.13})\,--\,(\ref{eq4.16}), the $\mathbb{Z}_2$ eigen mode function $\xi^{\eta} (z)$ takes zeros at the following fixed points \cite{Buchmuller:2015eya,Buchmuller:2018lkz}:
\begin{align}
\xi^{\eta}(p_1)&=0 
\qquad \textrm{for} \quad \eta=-1, \quad (\alpha_1,\alpha_2)=(0,0),(1/2,0),(0,1/2),(1/2,1/2),
\label{eq4.17}\\
\xi^{\eta}(p_2)&=0 \qquad \textrm{for} \quad
\begin{cases}
&\eta=1,\quad (\alpha_1,\alpha_2)=(1/2,0),(1/2,1/2),\\ 
&\textrm{or} \quad \eta=-1, \quad(\alpha_1,\alpha_2)=(0,0),(0,1/2),
\end{cases}
\label{eq4.18}\\
\xi^{\eta}(p_3)&=0 \qquad \textrm{for} \quad
\begin{cases}
&\eta=1,\quad (\alpha_1,\alpha_2)=(0,1/2),(1/2,1/2), \\
&\textrm{or} \quad \eta=-1, \quad(\alpha_1,\alpha_2)=(0,0),(1/2,0),
\end{cases}
\label{eq4.19}\\
\xi^{\eta}(p_4)&=0 \qquad \textrm{for} \quad
\begin{cases}
&\eta=1,\quad (\alpha_1,\alpha_2)=(1/2,0),(0,1/2),\\
&\textrm{or} \quad\eta=-1, \quad(\alpha_1,\alpha_2)=(0,0),(1/2,1/2).
\end{cases}
\label{eq4.20}
\end{align}
These relations imply that the winding numbers at the fixed points become nontrivial for the above zeros of the $\mathbb{Z}_2$ eigen mode function  $\xi^{\eta} (z)$. 
From Eqs.~\eqref{eq4.13}\,--\,(\ref{eq4.16})  we can compute the winding numbers $\chi_{\pm j}$ around the fixed points $p_j $. The results are summarized in Table \ref{Z2windingtab}.
From Tables \ref{Z2indextab} and \ref{Z2windingtab}, we can see that the 
coefficient $W_{j}$ in Eq.\,\eqref{eq5.12} is related to the winding numbers
$\chi_{\pm j}$ as
\begin{gather}
W_{j} = -\chi_{+j} + \chi_{-j} \qquad (j=1,2,3,4)
\label{neweq4.20.1}
\end{gather}
for any $\eta=\pm1$ and $(\alpha_1,\alpha_2)=(0,0),(1/2,0),(0,1/2)$ and $(1/2,1/2)$.
This implies that the index formula on $T^{2}/\mathbb{Z}_{2}$ is given by
\begin{gather}
n_{+} - n_{-} = \frac{1}{4}\,(-V_{+} + V_{-}),
\label{neweq4.20.2}
\end{gather}
where $V_{\pm}$ are the sums of the winding numbers $\chi_{\pm j}$ at the
fixed points $p_{j}\ (\text{or}\ \bm{y}^{f}_{j})$, i.e.
\begin{gather}
V_{\pm} = \sum_{j}\,\chi_{\pm j}.
\label{neweq4.20.3}
\end{gather}

\begin{table}[!t]
	\centering
	{\tabcolsep = 3.0mm
		\renewcommand{\arraystretch}{1.2}
		\scalebox{0.85}{
			\begin{tabular}{cc|cccc:cccc|c:c|c} \hline
				$\mathbb{Z}_2$ & twist &\multicolumn{8}{c|}{winding number}  &\multicolumn{2}{c|}{$\sum_j \chi_{\pm j}$}&index \\
				$\eta$  & $(\alpha_1, \alpha_2)$ & $\chi_{+1}$ & $\chi_{+2}$ &$\chi_{+3}$ &$\chi_{+4}$ &$\chi_{-1}$ &$\chi_{-2}$ &$\chi_{-3}$ &$\chi_{-4}$ &$V_+$ & $V_-$ &$(-V_+ + V_-)/4$ \\ \hline
				$1$  & $(0,0)$ & $0$ &$0$ &$0$ &$0$ &$1$ &$1$ &$1$ &$1$ & $0$ & $4$ & $1$ \\
				& $(1/2,0)$ & $0$ &$1$ &$0$ &$1$ &$1$ &$0$ &$1$ &$0$ & $2$ & $2$ & $0$ \\
				& $(0,1/2)$ & $0$ &$0$ &$1$ &$1$ &$1$ &$1$ &$0$ &$0$ & $2$ & $2$ & $0$ \\
				& $(1/2,1/2)$ & $0$ &$1$ &$1$ &$0$ &$1$ &$0$ &$0$ &$1$ & $2$ & $2$ & $0$ \\\cline{1-13}
				$-1$  & $(0,0)$ & $1$ &$1$ &$1$ &$1$ &$0$ &$0$ &$0$ &$0$ & $4$ & $0$ & $-1$ \\
				& $(1/2,0)$ & $1$ &$0$ &$1$ &$0$ &$0$ &$1$ &$0$ &$1$ & $2$ & $2$ & $0$ \\
				& $(0,1/2)$ & $1$ &$1$ &$0$ &$0$ &$0$ &$0$ &$1$ &$1$ & $2$ & $2$ & $0$ \\
				& $(1/2,1/2)$ & $1$ &$0$ &$0$ &$1$ &$0$ &$1$ &$1$ &$0$ & $2$ & $2$ & $0$ \\\hline
			\end{tabular}
		}
	}	
	\caption{The winding numbers $\chi_{\pm j}$ at the fixed points $p_j$ on $T^2/\mathbb{Z}_2$ and their sums $V_{\pm}=\sum_j \chi_{\pm j}$. All the values of $(-V_+ + V_-)/4$ exactly agree with the index $n_+ - n_-$ for the chiral zero modes.}
	\label{Z2windingtab}
\end{table}

\subsubsection*{\underline{B. Winding numbers for $T^2/\mathbb{Z}_3$}}
%

In the following, we examine the winding numbers $\chi_{\pm j}$ for the mode functions $f_+ (z) = \xi^{\eta} (z)$ and  $f_-(z) = \xi^{\omega \eta} (z)$ on $T^2/\mathbb{Z}_3$ at the fixed points \eqref{eq5.21}, which correspond to
\begin{gather}
p_1=0,\quad p_2=\frac{2}{3}+\frac{\tau}{3},\quad p_3=\frac{1}{3}+\frac{2\tau}{3}.\,
\label{eq4.21}
\end{gather}
in the complex plane.

Under the $\mathbb{Z}_3$  rotation $z \, \to \, \omega z \,\,(\omega=e^{i2\pi/3})$ around the fixed points $p_j$, the $\mathbb{Z}_3$ eigen mode function $\xi^{\eta}(z)$ is found to satisfy the relations 
\begin{align}
\xi^{\eta}(\omega z)&=\omega^{k} \xi^{\eta} (z),
\label{eq4.22}\\
\xi^{\eta}(\omega z+{2}/{3}+{\tau}/{3})&=\omega^{k+3\alpha_1} 
\xi^{\eta}(z+{2}/{3}+{\tau}/{3}),
\label{eq4.23}\\
\xi^{\eta}(\omega z+{1}/{3}+{2\tau}/{3})&=\omega^{k+3\alpha_1+3\alpha_2} 
\xi^{\eta} (z+{1}/{3}+{2\tau}/{3}),
\label{eq4.24}
\end{align}
where $\eta=\omega^k \,\,(k=0,1,2)$.

The relations (\ref{eq4.22})\,--\,(\ref{eq4.24}) tell the phase shifts to the $\mathbb{Z}_3$ eigen mode function $\xi^{\eta} (z)$ when rotated by $2\pi/3$ around each fixed point. To evaluate the winding number $\chi_j (\eta,\bm{\alpha})$ around the fixed point $p_j$, all we should do is to utilize the relations (\ref{eq4.22})\,--\,(\ref{eq4.24}) three times repeatedly. Then, we obtain
\begin{align}
\chi_1 (\eta,\bm{\alpha}) &= k \quad \textrm{mod}\, 3,
\label{eq4.25}\\
\chi_2 (\eta,\bm{\alpha}) &= k+3\alpha_1 \quad \textrm{mod} \,3,
\label{eq4.26}\\
\chi_3 (\eta,\bm{\alpha}) &= k+3\alpha_1+3\alpha_2 \quad \textrm{mod} \,3,
\label{eq4.27}
\end{align}
where $\eta=\omega^k$.  
From Eqs.~\eqref{eq4.25}\,--\,(\ref{eq4.27})  the winding numbers $\chi_{+j}=\chi_j (\eta,\bm{\alpha})$ and $\chi_{-j}=\chi_j (\bar{\omega}\bar{\eta},-\bm{\alpha})$ can be read off and are summarized in Table \ref{Z3windingtab}. 
From Tables \ref{Z3indextab} and \ref{Z3windingtab}, we can see that the relation
\begin{gather}
W_{j} = -\chi_{+j} + \chi_{-j} \qquad (j=1,2,3)
\label{neweq4.27.1}
\end{gather}
holds for any $\eta=1, \omega, \omega^{2}$ and 
$(\alpha_1,\alpha_2)=(0,0),(1/3,1/3)$ and $(2/3,2/3)$.
Thus, the index formula on $T^{2}/\mathbb{Z}_{3}$ is found to be 
\begin{gather}
n_{+} - n_{-} = \frac{1}{6}\,(-V_{+} + V_{-}).
\label{neweq4.27.2}
\end{gather}
%

\begin{table}[!t]
	\centering
	{\tabcolsep = 4.5mm
		\renewcommand{\arraystretch}{1.2}
		\scalebox{0.85}{
			\begin{tabular}{cc|ccc:ccc|c:c|c} \hline
				$\mathbb{Z}_3$ & twist &\multicolumn{6}{c|}{winding number}  &\multicolumn{2}{c|}{$\sum_j \chi_{\pm j}$}&index \\
				$\eta$  & $(\alpha_1, \alpha_2)$ & $\chi_{+1}$ & $\chi_{+2}$ &$\chi_{+3}$  &$\chi_{-1}$ &$\chi_{-2}$ &$\chi_{-3}$  &$V_+$ & $V_-$ &$(-V_+ + V_-)/6$ \\ \hline
				$1$  & $(0,0)$ & $0$ &$0$ &$0$ &$2$ &$2$ &$2$ & $0$ & $6$ & $1$ \\
				& $(1/3,1/3)$ & $0$ &$1$ &$2$ &$2$ &$1$ &$0$  & $3$ & $3$ & $0$ \\
				& $(2/3,2/3)$ & $0$ &$2$ &$1$ &$2$ &$0$ &$1$ &$3$ &$3$ &  $0$ \\\cline{1-11}
				$\omega$  & $(0,0)$ & $1$ &$1$ &$1$ &$1$ &$1$ &$1$ & $3$ & $3$ & $0$ \\
				& $(1/3,1/3)$ & $1$ &$2$ &$0$ &$1$ &$0$ &$2$  & $3$ & $3$ & $0$ \\
				& $(2/3,2/3)$ & $1$ &$0$ &$2$ &$1$ &$2$ &$0$ &$3$ &$3$ &  $0$ \\\cline{1-11}
				$\omega^2$  & $(0,0)$ & $2$ &$2$ &$2$ &$0$ &$0$ &$0$ & $6$ & $0$ & $-1$ \\
				& $(1/3,1/3)$ & $2$ &$0$ &$1$ &$0$ &$2$ &$1$  & $3$ & $3$ & $0$ \\
				& $(2/3,2/3)$ & $2$ &$1$ &$0$ &$0$ &$1$ &$2$ &$3$ &$3$ &  $0$ \\\hline
			\end{tabular}
		}
	}	
	\caption{The winding numbers $\chi_{\pm j}$ at the fixed points $p_j$ on $T^2/\mathbb{Z}_3$ and their sums $V_{\pm}=\sum_j \chi_{\pm j}$. All the values of $(-V_+ + V_-)/6$ exactly agree with the index $n_+ - n_-$ for the chiral zero modes.}
	\label{Z3windingtab}
\end{table}

We would like to make two comments on the winding numbers $\chi_{\pm j}$. As found from Table \ref{Z3windingtab}, $\chi_{\pm j}$ are equal to $0,1$ or $2$, as discussed in \cite{Sakamoto:2020pev}. The second comment is that we here consider the fundamental domain of $T^2$ in order to define the winding numbers  $\chi_{\pm j} $. We have defined the winding number $\chi_j (\eta,\bm{\alpha})$ in Eq.\,\eqref{eq4.3}, where the contour $C_j$ is taken to be a circle encircling the fixed point $p_j$. If the winding number is defined on the fundamental domain of $T^2/\mathbb{Z}_3$, it should be divided by $N=3$ due to deficit angles around the fixed points.

\subsubsection*{\underline{C. Winding numbers for $T^2/\mathbb{Z}_4$}}
%

In the following, we examine the winding numbers $\chi_{\pm j}$ for the mode functions $f_+ (z) = \xi^{\eta} (z)$ and  $f_-(z) = \xi^{\omega \eta} (z)$ on $T^2/\mathbb{Z}_4$ at the fixed points.

As noted in the previous section, there are two $\mathbb{Z}_{4}$ fixed points
\begin{gather}
p_1=0,\quad p_2=\frac{1}{2}+\frac{\tau}{2},
\label{eq4.28}
\end{gather}
and additionally two ``$\mathbb{Z}_2$ fixed points"
\begin{gather}
\mathbb{Z}_2\, \textrm{fixed points}:\,p_3=\frac{1}{2},\quad p_4=\frac{\tau}{2},
\label{eq4.29}
\end{gather}
which are not invariant under the $\mathbb{Z}_{4}$ rotation but invariant
under the $\mathbb{Z}_{2}$ one $(z \to \omega^2 z = -z)$ up to torus lattice shifts.
The winding numbers not only at the $\mathbb{Z}_4$ fixed points (\ref{eq4.28}) 
but also at the ``$\mathbb{Z}_2$ fixed points" (\ref{eq4.29}) contribute to 
the formula (\ref{eq5.25}).

%
%

Under the $\mathbb{Z}_4$  rotation $z \, \to \, \omega z $ around the fixed points $p_1$ and $p_2$, and under the $\mathbb{Z}_2$  rotation $z \, \to \, \omega^2 z $ around the fixed points $p_3$ and $p_4$, the $\mathbb{Z}_4$ eigen mode function $\xi^{\eta}(z)$ is found to satisfy the relations 
\begin{align}
\xi^{\eta}(\omega z)&=\omega^{k} \xi^{\eta} (z),
\label{eq4.30}\\
\xi^{\eta}(\omega z+{1}/{2}+{\tau}/{2})
  &=\omega^{k+4\alpha_1} \xi^{\eta} (z+{1}/{2}+{\tau}/{2}),
\label{eq4.31}\\
\xi^{\eta}(\omega^2 z+{1}/{2})&=(\omega^2)^{k+2\alpha_1} \xi^{\eta} (z+{1}/{2}),
\label{eq4.32}\\
\xi^{\eta}(\omega^2 z+{\tau}/{2})&=(\omega^2)^{k+2\alpha_2} \xi^{\eta} (z+{\tau}/{2}),
\label{eq4.33}
\end{align}
where $\eta=\omega^k \,\,(k=0,1,2,3)$.

From Eqs.~\eqref{eq4.30}\,--\,(\ref{eq4.33}) the winding number $\chi_j (\eta,\bm{\alpha})$ around the fixed point $p_j$ is found as
\begin{align}
\chi_1 (\eta,\bm{\alpha}) &= k \qquad \textrm{mod} \,\,4,
\label{eq4.34}\\
\chi_2 (\eta,\bm{\alpha}) &= k+4\alpha_1 \qquad \textrm{mod} \,\,4,
\label{eq4.35}\\
\chi_3 (\eta,\bm{\alpha}) &= k+2\alpha_1 \qquad \textrm{mod} \,\,2,
\label{eq4.36}\\
\chi_4 (\eta,\bm{\alpha}) &= k+2\alpha_2 \qquad \textrm{mod} \,\,2,
\label{eq4.37}
\end{align}
where $\eta=\omega^k$. Eqs.~\eqref{eq4.34}\,--\,(\ref{eq4.37}) show that the winding numbers $\chi_{\pm j}$ can be read off and are summarized in Table \ref{Z4windingtab}. 
From Tables \ref{Z4indextab} and \ref{Z4windingtab}, we can see that the relation
\begin{gather}
W_{j} = -\chi_{+j} + \chi_{-j} \qquad (j=1,2,3,4)
\label{neweq4.37.1}
\end{gather}
holds for any $\eta=\omega^{k}\ (k=0,1,2,3)$ and 
$(\alpha_1,\alpha_2)=(0,0)$ and $(1/2,1/2)$.
Thus, we find the index formula on $T^{2}/\mathbb{Z}_{4}$ as
\begin{gather}
n_{+} - n_{-} = \frac{1}{8}\,(-V_{+} + V_{-}).
\label{neweq4.37.2}
\end{gather}
As one can see from Table \ref{Z4windingtab}, $\chi_{\pm 1} $ and $\chi_{\pm 2} $ ($\chi_{\pm 3} $ and $\chi_{\pm 4} $) are equal to $0,1,2,3 \,(0,1)$, as discussed in \cite{Sakamoto:2020pev}. In fact,  the numbers in Table \ref{Z4windingtab} lead to the formula  (\ref{neweq4.37.2}).
%

\begin{table}[!t]
	\centering
	{\tabcolsep = 3.0mm
		\renewcommand{\arraystretch}{1.2}
		\scalebox{0.85}{
			\begin{tabular}{cc|cccc:cccc|c:c|c} \hline
				$\mathbb{Z}_4$ & twist &\multicolumn{8}{c|}{winding number}  &\multicolumn{2}{c|}{$\sum_j \chi_{\pm j}$}&index \\
				$\eta$  & $(\alpha_1, \alpha_2)$ & $\chi_{+1}$ & $\chi_{+2}$ &$\chi_{+3}$ &$\chi_{+4}$  &$\chi_{-1}$ &$\chi_{-2}$ &$\chi_{-3}$  &$\chi_{-4}$ &$V_+$ & $V_-$ &$(-V_+ + V_-)/8$ \\ \hline
				$1$  & $(0,0)$ & $0$ & $0$ &$0$ &$0$ &$3$ &$3$ &$1$ & $1$ & $0$ & $8$ &$1$ \\
				& $(1/2,1/2)$ & $0$ & $2$ &$1$ &$1$ &$3$ &$1$ &$0$ & $0$ & $4$ & $4$ &$0$ \\ \cline{1-13}
				$\omega$  & $(0,0)$ & $1$ & $1$ &$1$ &$1$ &$2$ &$2$ &$0$ & $0$ & $4$ & $4$ &$0$ \\
				& $(1/2,1/2)$ & $1$ & $3$ &$0$ &$0$ &$2$ &$0$ &$1$ & $1$ & $4$ & $4$ &$0$ \\ \cline{1-13}
				$\omega^2$  & $(0,0)$ & $2$ & $2$ &$0$ &$0$ &$1$ &$1$ &$1$ & $1$ & $4$ & $4$ &$0$ \\
				& $(1/2,1/2)$ & $2$ & $0$ &$1$ &$1$ &$1$ &$3$ &$0$ & $0$ & $4$ & $4$ &$0$ \\ \cline{1-13}
				$\omega^3$  & $(0,0)$ & $3$ & $3$ &$1$ &$1$ &$0$ &$0$ &$0$ & $0$ & $8$ & $0$ &$-1$ \\
				& $(1/2,1/2)$ & $3$ & $1$ &$0$ &$0$ &$0$ &$2$ &$1$ & $1$ & $4$ & $4$ &$0$ \\ \hline
			\end{tabular}
		}
	}	
	\caption{The winding numbers $\chi_{\pm j}$ at the fixed points $p_j$ on $T^2/\mathbb{Z}_4$ and their sums $V_{\pm}=\sum_j \chi_{\pm j}$. All the values of $(-V_+ + V_-)/8$ exactly agree with the index $n_+ - n_-$ for the chiral zero modes.}
	\label{Z4windingtab}
\end{table}

\subsubsection*{\underline{D. Winding numbers for $T^2/\mathbb{Z}_6$}}
%

In the following, we examine the winding numbers $\chi_{\pm j}$ for the mode functions $f_+ (z) = \xi^{\eta} (z)$ and  $f_-(z) = \xi^{\omega \eta} (z)$ on $T^2/\mathbb{Z}_6$ at the fixed points.

As noted in the previous section, there is only a single $\mathbb{Z}_{6}$ fixed point 
\begin{gather}
p_1=0,
\label{eq4.38}
\end{gather}
and there are additionally two ``$\mathbb{Z}_3$ fixed points" and 
three ``$\mathbb{Z}_2$ fixed points" given by
\begin{align}
\mathbb{Z}_3~\textrm{fixed points}:& \,~p_2=\frac{1}{3}+\frac{\tau}{3},\quad p_3=\frac{2}{3}+\frac{2\tau}{3},
\label{eq4.39}\\
\mathbb{Z}_2~\textrm{fixed points}:& \,~p_4=\frac{1}{2},\quad p_5=\frac{\tau}{2},\quad p_6=\frac{1}{2}+\frac{\tau}{2}.
\label{eq4.40}
\end{align}
The winding numbers not only at the $\mathbb{Z}_{6}$ fixed point \eqref{eq4.38}
but also at the ``$\mathbb{Z}_3$ fixed points" \eqref{eq4.39} 
and the ``$\mathbb{Z}_2$ fixed points" \eqref{eq4.40} contribute to the
formula \eqref{eq5.33}.


Under the $\mathbb{Z}_6$ ($\mathbb{Z}_3$ and $\mathbb{Z}_2$) rotation $z \, \to \, \omega z \,\,(z \, \to \, \omega^2 z \,\,$and$ \,\,z \, \to \, \omega^3 z)$ around the fixed points $p_1$ ($p_2$, $p_3$ and $p_4,p_5,p_6$), the $\mathbb{Z}_6$  eigen mode function $\xi^{\eta}(z)$ is found to satisfy the relations 
\begin{align}
\xi^{\eta}(\omega z)&=\omega^{k} \xi^{\eta} (z),
\label{eq4.41}\\
\xi^{\eta}(\omega^2 z+p_j)&=(\omega^2)^k \xi^{\eta} (z+p_j) \quad (j=2,3),
\label{eq4.42}\\
\xi^{\eta}(\omega^3 z+p_j)&=(\omega^3)^k \xi^{\eta} (z+p_j) \quad (j=4,5,6),
\label{eq4.43}
\end{align}
where $\eta=\omega^k \,\,(k=0,1,\cdots,5)$.

From Eqs.~\eqref{eq4.41}\,--\,(\ref{eq4.43}), the winding numbers $\chi_j \,(\eta,\bm{\alpha}=\bm 0)$ around the fixed points $p_j $ are found to satisfy
\begin{align}
\chi_1 (\eta,\bm{\alpha}=\bm 0)&= k \qquad \textrm{mod} \,\,6,
\label{eq4.44}\\
\chi_j (\eta,\bm{\alpha}=\bm 0) &= k \qquad \textrm{mod} \,\,3 \quad(j=2,3),
\label{eq4.45}\\
\chi_j (\eta,\bm{\alpha}=\bm 0) &= k \qquad \textrm{mod}\,\, 2\quad(j=4,5,6).
\label{eq4.46}
\end{align}
Note that the SS twist phase is restricted to $\bm{\alpha}=\bm 0$ on $T^2/\mathbb{Z}_6$. 
From Eqs.~\eqref{eq4.44}\,--\,(\ref{eq4.46}) the winding numbers $\chi_{\pm j}$ can be read off and are summarized in Table \ref{Z6windingtab}.
From Tables \ref{Z6indextab} and \ref{Z6windingtab}, 
we can see that the relation
\begin{gather}
W_{j} = -\chi_{+j} + \chi_{-j} \qquad (j=1,2,\cdots,6)
\label{neweq4.46.1}
\end{gather}
holds for any $\eta=\omega^{k}\ (k=0,1,\cdots,5)$ with $\bm{\alpha}=\bm 0$.
Thus, we find the index formula on $T^{2}/\mathbb{Z}_{6}$ as
\begin{gather}
n_{+} - n_{-} = \frac{1}{12}\,(-V_{+} + V_{-}).
\label{neweq4.46.2}
\end{gather}
%
%
As one can see from Table \ref{Z6windingtab}, $\chi_{\pm 1} $ ($\chi_{\pm 2} $, $\chi_{\pm 3} $ and $\chi_{\pm 4} \,$, $\chi_{\pm 5} \,$, $\chi_{\pm 6} $) are equal to $0,1,2,3,4,5$ \,($0$, $1$, $2$\, and $\,0,1$), as discussed in \cite{Sakamoto:2020pev}. In fact, the winding numbers in Table \ref{Z6windingtab} lead to the formula  (\ref{neweq4.46.2}).

\begin{table}[!t]
	\centering
	{\tabcolsep = 2.0mm
		\renewcommand{\arraystretch}{1.2}
		\scalebox{0.85}{
			\begin{tabular}{c|cccccc:cccccc|c:c|c} \hline
				$\mathbb{Z}_6$  &\multicolumn{12}{c|}{winding number}  &\multicolumn{2}{c|}{$\sum_j \chi_{\pm j}$}&index \\
				$\eta$  &  $\chi_{+1}$ & $\chi_{+2}$ &$\chi_{+3}$ &$\chi_{+4}$  &$\chi_{+5}$  &$\chi_{+6}$  &$\chi_{-1}$ &$\chi_{-2}$ &$\chi_{-3}$  &$\chi_{-4}$ &$\chi_{-5}$  &$\chi_{-6}$  &$V_+$ & $V_-$ &$(-V_+ + V_-)/12$ \\ \hline
				$1$  & $0$ & $0$ & $0$ &$0$ &$0$ &$0$ &$5$ &$2$ & $2$ & $1$ & $1$ &$1$& $0$ & $12$ &$1$ \\\hline
				$\omega$  & $1$ & $1$ & $1$ &$1$ &$1$ &$1$ &$4$ &$1$ & $1$ & $0$ & $0$ &$0$& $6$ & $6$ &$0$ \\\hline
				$\omega^2$  & $2$ & $2$ & $2$ &$0$ &$0$ &$0$ &$3$ &$0$ & $0$ & $1$ & $1$ &$1$& $6$ & $6$ &$0$ \\\hline
				$\omega^3$  & $3$ & $0$ & $0$ &$1$ &$1$ &$1$ &$2$ &$2$ & $2$ & $0$ & $0$ &$0$& $6$ & $6$ &$0$ \\\hline
				$\omega^4$  & $4$ & $1$ & $1$ &$0$ &$0$ &$0$ &$1$ &$1$ & $1$ & $1$ & $1$ &$1$& $6$ & $6$ &$0$ \\\hline
				$\omega^5$  & $5$ & $2$ & $2$ &$1$ &$1$ &$1$ &$0$ &$0$ & $0$ & $0$ & $0$ &$0$& $12$ & $0$ &$-1$ \\\hline
			\end{tabular}
		}
	}	
	\caption{The winding numbers $\chi_{\pm j}$ at the fixed points $p_j$ on $T^2/\mathbb{Z}_6$ and their sums $V_{\pm}=\sum_j \chi_{\pm j}$. Here, we omit the column of $(\alpha_1,\alpha_2)$ due to the fact that $(\alpha_1,\alpha_2)=(0,0)$ for the $T^2/\mathbb{Z}_6$ orbifold. All the values of $(-V_+ + V_-)/12$ exactly agree with the index $n_+ - n_-$ for the chiral zero modes.}
	\label{Z6windingtab}
\end{table}

\section{Conclusion and discussion}
%

In this paper, we have derived the index formula
\begin{gather}
n_+ -n_- =\frac{1}{2N} (-V_+  +V_-)
\label{eq6.1}
\end{gather}
on the $T^2/\mathbb{Z}_N$ orbifold from the trace formula \eqref{eq5.1}.
In Section 3, we have explicitly constructed the mode functions on $T^2/\mathbb{Z}_N$ 
and counted the numbers $n_{\pm}$ of the chiral zero modes.
In Sections 4 and 5, we have succeeded in evaluating the trace formula \eqref{eq5.1}
and clearly shown that the index $n_{+}-n_{-}$ is determined by the winding
numbers at the fixed points on $T^{2}/\mathbb{Z}_{N}$.


We have emphasized that the dependence of $n_{\pm}$ on $N$, $\eta$ and $(\alpha_1,\alpha_2)$ 
is rather simple, as shown in Tables \ref{Z2tab}\,--\,\ref{Z6tab}, 
but the equality in Eq.\,\eqref{eq6.1} is nontrivial. 
This is because the values of $V_{\pm}$ (or $\chi_{\pm j}$) can be non-vanishing even if $n_+$ and/or $n_-$ are zero, as seen in Tables \ref{Z2windingtab}\,--\,\ref{Z6windingtab}. 
Furthermore, $V_{\pm}/2N$ are not integer-valued in general, 
but the difference $(-V_{+}+V_-)/2N$ becomes an integer in any case.

It is interesting that from Tables \ref{Z2windingtab}\,--\,\ref{Z6windingtab} the sums $V_{\pm}$ of the winding numbers at the fixed points satisfy the relation
\begin{gather}
V_+ +V_-=2N,
\label{eq6.3}
\end{gather}
which may be regarded as an expression of the index theorem.
Then, from Eqs.~\eqref{eq6.1} and (\ref{eq6.3}) we have
\begin{gather}
  n_+ -n_-=-\frac{V_+}{N} +1.
  \label{eq6.4}
 \end{gather}
This can be understood as a special case of $M=0$ in the zero-mode counting formula \cite{Sakamoto:2020pev}
\begin{gather}
n_+ -n_-=\frac{M-V_+}{N} +1.
\label{eq6.5}
\end{gather}
There the formula (\ref{eq6.5}) with the quantized magnetic flux $M$ has been confirmed only for $M>0$.
Since we can show that the relation (\ref{eq6.3}) holds also for $M\neq 0$, the formula (\ref{eq6.5}) can be rewritten as 
\begin{gather}
n_+ -n_-=\frac{1}{2N} (2M-V_+ +V_-),
\label{eq6.6}
\end{gather}
which leads to a generalization of the formula (\ref{eq6.1}) to $M\neq 0$.

In this paper, we found that the orbifold projections bring the chirality of the massless level in the KK tower from the vewpoint of the index theorem (even if no flux is turned on).
In addition, the orbifold fixed points and zero points there play important roles in the trace theorem, as we expected from the previous paper \cite{Sakamoto:2020pev}.
Thus, these evidences let us conclude that the term $- V_+/N +1$ or $(-V_+ + V_-)/2N$ reflects the contribution from the orbifold geometry, i.e. the singularity of the fixed points.

There are two possibilities to interpret the term $- V_+/N +1$ or $(-V_+ + V_-)/2N$.
One is that the term originates from some singular spin connection or curvature at the orbifold fixed points, which should be regarded as ``geometric flux''.
Another possibility is that there exist localized Wilson-line sources at the fixed points, as discussed in \cite{Buchmuller:2015eya, Buchmuller:2018lkz}.
In this case such sources should be regarded as ``gauge flux''.
Our result suggests that the two-dimensional orbifolds $T^2/\mathbb{Z}_N$ are equivalently described by setups with localized fluxes at fixed points.

A remaining task is to derive the formula (\ref{eq6.6}) for $M\neq 0$ from the trace formula. 
In order to evaluate the trace formula, we may need a complete orthonormal set
of $\mathbb{Z}_{N}$ eigen mode functions on
the magnetized $T^{2}/\mathbb{Z}_{N}$ orbifold, as we have done in this paper.
That, however, seems to be hard since mode functions on the magnetized $T^{2}/\mathbb{Z}_{N}$
are given by Jacobi theta functions \cite{Cremades:2004wa} and their $\mathbb{Z}_{N}$ 
transformation property is quite complicated 
\cite{Abe:2013bca,Abe:2014noa}.



\section*{Acknowledgment}
Y.T. would like to thank Wilfried Buchm\"uller and Markus Dierigl for instructive comments on this manuscript.
This work is supported by the Deutsche Forschungsgemeinschaft under Germany's Excellence Strategy - EXC 2121 Quantum Universe - 390833306.
M.S. is supported by Japan Society for the Promotion of Science (JSPS) KAKENHI Grant Number JP\,18K03649.
Y.T. is supported in part by Grants-in-Aid for JSPS Overseas Research Fellow (No.\,18J60383) from the Ministry of Education, Culture, Sports, Science and Technology in Japan, and in part by Scuola Normale, by INFN (IS GSS-Pi) and by the MIUR-PRIN contract 2017CC72MK\_003.

\appendix
\renewcommand{\thesection}{\Alph{section}}

\section{Appendix}

In this appendix, we prove the relation 
\begin{align}
\sum_{\bm n +\bm{\alpha}\, \in \Lambda / {\mathbb{Z}_{N}} } &\{ \xi_{\bm n +\bm{\alpha}}^{\eta} (z)(\xi_{\bm n +\bm{\alpha}}^{\eta} (z'))^{\ast}-  \xi_{\bm n +\bm{\alpha}}^{\omega \eta} (z)(\xi_{\bm n +\bm{\alpha}}^{\omega \eta} (z'))^{\ast}\} \notag \\
&=\frac{1}{\textrm{Im} \tau}\sum_{n_1 \in \mathbb{Z}} \sum_{n_2 \in \mathbb{Z}} \sum_{l=0}^{N-1} \eta^l (1-\omega^l) u_{\bm n +\bm{\alpha}}(z)(u_{\bm n +\bm{\alpha}} (\omega^l z'))^{\ast},
\label{eqA.1}
\end{align}
which is used in deriving Eq.\,\eqref{eq5.5}. To prove Eq.\,\eqref{eqA.1}, we separately discuss the following four cases:
\begin{align}
&(1) \quad \bm{\alpha} \neq \bm 0, \notag \\
&(2) \quad \bm{\alpha} = \bm 0 \quad \textrm{and} \quad \eta \neq 1, \quad \omega \eta \neq 1, \notag \\
&(3) \quad \bm{\alpha} = \bm 0 \quad \textrm{and} \quad \eta = 1, \quad \omega \eta \neq 1, \notag \\
&(4) \quad \bm{\alpha} = \bm 0 \quad \textrm{and} \quad \eta \neq 1, \quad \omega \eta = 1. \notag
\end{align}

\subsubsection*{\underline{$(1)~~ \bm{\alpha} \neq \bm 0$}}

For $\bm{\alpha} \neq \bm 0$, the left-hand-side of Eq.\,\eqref{eqA.1} can be evaluated as follows:
\begin{align}
I(z,z') \,
&\equiv \sum_{{\bm n}+\bm{\alpha}\, \in \Lambda / {\mathbb{Z}_{N}} } \{ \xi_{\bm n +\bm{\alpha}}^{\eta} (z)(\xi_{\bm n +\bm{\alpha}}^{\eta} (z'))^{\ast}-  \xi_{\bm n +\bm{\alpha}}^{\omega \eta} (z)(\xi_{\bm n +\bm{\alpha}}^{\omega \eta} (z'))^{\ast}\} \notag \\
&\overset{(\ref{eq3.5})}{=} 
\sum_{\bm n +\bm{\alpha}\, \in \Lambda / {\mathbb{Z}_{N}} } |A_{\bm n +\bm{\alpha}} |^2  \sum_{l=0}^{N-1}\sum_{l'=0}^{N-1} \eta^{-l+l'} (1-\omega^{-l+l'}) u_{\bm n +\bm{\alpha}}(\omega^l z)(u_{\bm n +\bm{\alpha}} (\omega^{l'} z'))^{\ast}\notag \\
&\overset{l'=l''+l}{=} \sum_{\bm n +\bm{\alpha}\, \in \Lambda / {\mathbb{Z}_{N}} } |A_{\bm n +\bm{\alpha}} |^2  \sum_{l=0}^{N-1}\sum_{l''=0}^{N-1} \eta^{l''} (1-\omega^{l''}) u_{\bm n +\bm{\alpha}}(\omega^{l}z)(u_{\bm n +\bm{\alpha}} (\omega^{l''+l} z'))^{\ast}\notag \\
&=\frac{1}{\textrm{Im} \tau}\sum_{n_1 \in \mathbb{Z}} \sum_{n_2 \in \mathbb{Z}} \sum_{l''=0}^{N-1} \eta^{l''} (1-\omega^{l''}) u_{\bm n +\bm{\alpha}}(z)(u_{\bm n +\bm{\alpha}} (\omega^{l''} z'))^{\ast},
\label{eqA.2}
\end{align}
where in the last equality, we have used Eqs.~\eqref{eq3.11}, (\ref{eq3.18})  and the formula 
\begin{align}
 \notag
\sum_{\bm n +\bm{\alpha}\, \in \Lambda / {\mathbb{Z}_{N}} }\sum_{l=0}^{N-1} F(\omega^l (\bm n +\bm{\alpha}))&= \sum_{\bm n +\bm{\alpha}\, \in \Lambda  } F(\bm n +\bm{\alpha}) \notag \\
&=\sum_{n_1 \in \mathbb{Z}} \sum_{n_2 \in \mathbb{Z}} 
 F(\bm n +\bm{\alpha})
\label{eqA.3}
\end{align}
for $\bm{\alpha} \neq \bm 0$.

\subsubsection*{\underline{$(2) ~~ \bm{\alpha} = \bm 0 ~~ \textrm{and} ~~\eta \neq 1, ~ \omega \eta \neq 1$}}

In this case, $\Lambda / {\mathbb{Z}_{N}}$ should not include ${\bm n}=\bm0$ (see Eq.\,\eqref{eq3.15}). Hence, we can write $\Lambda / {\mathbb{Z}_{N}}$ explicitly as
\begin{gather}
\Lambda^{'} / {\mathbb{Z}_{N}} \equiv \left\{ \bm{n} ~(n_1,n_2 \in  {\mathbb{Z}}) ~~ | ~~\bm{n} \,\, \sim \omega\bm{n} ~~ \textrm{and} ~~\bm{n}\neq \bm 0 \right\}
\label{eqA.4}
\end{gather}
and
\begin{gather}
\Lambda^{'} \equiv \left\{  \bm{n} ~(n_1,n_2 \in  {\mathbb{Z}}) ~~ | ~~ \bm{n}\neq \bm 0 \right\}.
\label{eqA.5}
\end{gather}
Then, we find 
 \begin{align}
 I(z,z') 
 &\equiv \sum_{\bm n \, \in \Lambda^{'} / {\mathbb{Z}_{N}} } \{ \xi_{\bm n }^{\eta} (z)(\xi_{\bm n }^{\eta} (z'))^{\ast}-  \xi_{\bm n }^{\omega \eta} (z)(\xi_{\bm n }^{\omega \eta} (z'))^{\ast}\} \notag \\
 &\overset{(\ref{eq3.5})}{=} 
 \sum_{\bm n \, \in \Lambda^{'} / {\mathbb{Z}_{N}} } |A_{\bm n} |^2  \sum_{l=0}^{N-1}\sum_{l'=0}^{N-1} \eta^{-l+l'} (1-\omega^{-l+l'}) u_{\bm n }(\omega^l z)(u_{\bm n} (\omega^{l'} z'))^{\ast}\notag \\
 &\overset{l'=l''+l}{=}\frac{1}{\rm{Im} \tau} \sum_{\bm n  \in \Lambda^{'} } \sum_{l''=0}^{N-1} \eta^{l''} (1-\omega^{l''}) u_{\bm n }(z)(u_{\bm n } (\omega^{l''} z'))^{\ast}\notag \\
 & \qquad \left(\because (\ref{eq3.11}), (\ref{eq3.18}) ~~ \textrm{and}  \sum_{\bm n \, \in \Lambda^{'} / {\mathbb{Z}_{N}} } \sum_{l=0}^{N-1} F(\omega^l \bm n )= \sum_{\bm n \, \in \Lambda^{'} }F(\bm n)\right) \notag \\
 &=\frac{1}{\textrm{Im} \tau} \sum_{n_1 \in \mathbb{Z}}\sum_{n_2 \in \mathbb{Z}} \sum_{l''=0}^{N-1} \eta^{l''} (1-\omega^{l''}) u_{\bm n }(z)(u_{\bm n } (\omega^{l''} z'))^{\ast},
 \label{eqA.6}
 \end{align}
 where in the last equality, we have used $u_{\bm 0} (z)=1$ and 
\begin{gather}
\sum_{l''=0}^{N-1} \eta^{l''}=\sum_{l''=0}^{N-1} (\omega \eta)^{l''}=0,
\label{eqA.7}
\end{gather} 
for $\eta \neq 1$ and $\omega \eta\neq1$.

 \subsubsection*{\underline{$(3) ~~ \bm{\alpha} = \bm 0 ~~ \textrm{and} ~~\eta = 1, ~\omega \eta \neq 1$}}

 In this case, $I(z,z')$ is expressed in terms of $\Lambda / {\mathbb{Z}_{N}}$ and $\Lambda^{'} / {\mathbb{Z}_{N}}$ as
 \begin{align}
 I(z,z')
 &= \sum_{\bm n \, \in \Lambda / {\mathbb{Z}_{N}} }  \xi_{\bm n }^{\eta} (z)(\xi_{\bm n }^{\eta} (z'))^{\ast}-  \sum_{\bm n \, \in \Lambda^{'} / {\mathbb{Z}_{N}} } \xi_{\bm n }^{\omega \eta} (z)(\xi_{\bm n }^{\omega \eta} (z'))^{\ast} \notag \\
 &\overset{(\ref{eq3.5})}{=} 
 \sum_{\bm n \, \in \Lambda^{'} / {\mathbb{Z}_{N}} } |A_{\bm n} |^2  \sum_{l=0}^{N-1}\sum_{l'=0}^{N-1}  (1-\omega^{-l+l'}) u_{\bm n }(\omega^l z)(u_{\bm n} (\omega^{l'} z'))^{\ast}+\frac{N}{\rm{Im} \tau} \notag \\
 &\qquad \left(\because \eta=1, \, |A_{\bm 0}|^2 =(N \textrm{Im} \tau)^{-1} ,\, u_{\bm 0} (z)=1 \,\right) \notag \\
 &=\frac{1}{\textrm{Im} \tau} \sum_{\bm n \in \Lambda^{'}} \sum_{l''=0}^{N-1} (1-\omega^{l''}) u_{\bm n }(z)(u_{\bm n } (\omega^{l''} z'))^{\ast}+\frac{N}{\rm{Im} \tau} \notag \\
 & \qquad \left(\because (\ref{eq3.11}), (\ref{eq3.18}) ~~ \textrm{and}  \sum_{\bm n \, \in \Lambda^{'} / {\mathbb{Z}_{N}} } \sum_{l=0}^{N-1} F(\omega^{l} \bm n )= \sum_{\bm n \, \in \Lambda^{'} }F(\bm n)\right) \notag \\
  &=\frac{1}{\textrm{Im} \tau} \sum_{\bm n \in \Lambda} \sum_{l''=0}^{N-1} (1-\omega^{l''}) u_{\bm n }(z)(u_{\bm n } (\omega^{l''} z'))^{\ast} \notag \\
   &\qquad \left(\because u_{\bm 0} (z)=1, \,\sum_{l''=0}^{N-1} (1-\omega^{l''}) =N \right) \notag \\
    &=\frac{1}{\textrm{Im} \tau} \sum_{ n_1 \in \mathbb{Z}}
     \sum_{ n_2 \in \mathbb{Z}} \sum_{l''=0}^{N-1} (1-\omega^{l''}) u_{\bm n }(z)(u_{\bm n } (\omega^{l''} z'))^{\ast} .
 \label{eqA.8}
 \end{align}

 \subsubsection*{\underline{$(4) ~~ \bm{\alpha} = \bm 0 ~~ \textrm{and} ~~\eta \neq 1, ~ \omega \eta = 1$}}

  We can prove the relation (\ref{eqA.1}) in the same way as the case $(3)$ for $\bm{\alpha} = \bm 0 ~~ \textrm{and} ~~\eta \neq 1, ~ \omega \eta = 1$.

\bibliography{references222}
\bibliographystyle{JHEP}

\end{document}